\documentclass[]{aa}

\usepackage{txfonts}
\usepackage{graphicx}
\usepackage{rotating}
\usepackage{natbib}
\bibpunct{(}{)}{;}{a}{}{,} 
\newcommand{\ha}{H$\alpha$}
\newcommand{\brg}{Br$\gamma$}
\newcommand{\pab}{Pa$\beta$}
\newcommand{\htwo}{H$_2$}
\newcommand{\hi}{\ion{H}{i}}
\newcommand{\nai}{\ion{Na}{i}}
\newcommand{\feii}{[\ion{Fe}{ii}]}
\newcommand{\sii}{[\ion{S}{ii}]}
\newcommand{\hei}{\ion{He}{i}}
\newcommand{\caii}{\ion{Ca}{ii}}
\newcommand{\oi}{[\ion{O}{i}]}

\newcommand{\um}{$\mu$m}

\newcommand{\lsun}{L$_{\odot}$}
\newcommand{\msun}{M$_{\odot}$}
\newcommand{\msunyr}{M$_{\odot}$\,yr$^{-1}$}
\newcommand{\macc}{$\dot{M}_{acc}$}
\newcommand{\mloss}{$\dot{M}_{loss}$}

\newcommand{\lacc}{$L_{\mathrm{acc}}$}
\newcommand{\rstar}{$R_{\mathrm{*}}$}
\newcommand{\lstar}{$L_{\mathrm{*}}$}
\newcommand{\mstar}{$M_{\mathrm{*}}$}

\begin{document}

\title{POISSON project -- I -- Emission lines as accretion tracers in young stellar objects: results from observations of Chamaeleon I and II sources\thanks{Based on observations collected at the European Southern 
Observatory, La Silla, Chile (ESO programme 082.C-0264).}}

\author{S. Antoniucci\inst{1},
		R. Garc\'ia L\'opez\inst{1,2},
		B. Nisini\inst{1},
		T. Giannini\inst{1},
		D. Lorenzetti\inst{1}
		J. Eisl\"offel\inst{3},
		F. Bacciotti\inst{4},
		S. Cabrit\inst{5},
		A. Caratti o Garatti\inst{6},
		C. Dougados\inst{7}
		\and T. Ray\inst{6},
		}

\institute{ INAF-Osservatorio Astronomico di Roma, Via di Frascati 33, 00040 Monte Porzio Catone, Italy \and
			Max-Planck-Institut f\"ur Radioastronomie, Auf dem H\"ugel 69,53121 Bonn, Germany \and
			Th\"uringer Landessternwarte Tautenburg, Sternwarte 5, D-07778 Tautenburg, Germany \and
			INAF-Osservatorio Astrofisico di Arcetri, Largo Enrico Fermi 5, 50125 Firenze, Italy \and
			Laboratoire d'Etudes du Rayonnement et de la Matière en Astrophysique (LERMA), Observatoire de Paris, ENS, UPMC, UCP, CNRS; 61 Avenue de l'Observatoire, 75014 Paris, France \and
			Dublin Institute for Advanced Studies, School of Cosmic Physics, 31 Fitzwilliam Place, Dublin 2, Ireland \and 
			Institut de Planetologie et d'Astrophysique de Grenoble (IPAG), UMR 5571, BP 53, 38041 Grenoble Cedex 09, France} 

\offprints{Simone Antoniucci, \email{simone.antoniucci@oa-roma.astro.it}}
\date{Received date / Accepted date}
\titlerunning{POISSON observations: emission lines as accretion tracers in YSOs}
\authorrunning{Antoniucci et al.}

\abstract
{We present the results of the analysis of low-resolution optical-near IR spectroscopy (0.6-2.4 \um) of a sample (47 sources) of Class I and Class II 
young stellar objects in the Chamaeleon I and II star-forming clouds. These data are part of the POISSON project (Protostellar Optical-Infrared 
Spectral Survey on NTT).}
{The aim of the observations is to determine the accretion luminosity (\lacc) and mass accretion rate (\macc) of the sources through the analysis of the detected emission 
features. Taking advantage of the wide wavelength range covered by our spectra, we also aim at verifying 
the reliability and consistency of the existing empirical relationships connecting emission line luminosity and \lacc.}
{We employ five different tracers (\oi\,$\lambda6300$, \ha, \caii\,$\lambda8542$, \pab, and \brg) 
to derive the accretion luminosity, and critically discuss the various determinations in the light of the source properties.}
{The tracers provide \lacc\, values characterised by different scatters when plotted as a function of \lstar.
The \brg\, relation appears to be the most reliable, because it gives the minimum dispersion of \lacc\, over the entire range of \lstar, whereas the 
other tracers, in particular \ha, provide much more scattered \lacc\, results, which are not expected for the homogeneous sample of targets we are observing.
The direct comparison between \lacc(\brg) and the accretion luminosity obtained from the other four tracers also shows systematic differences 
in the results provided by the empirical relationships. These may probably be ascribed to different 
excitation mechanisms that contribute to the line emission, which may vary between our sample and those where the relationships have been calibrated,
which were mostly based on observations in Taurus.
Adopting the accretion luminosities estimates derived from the \brg\, line, we infer \lacc\, in the range 0.1~\lstar--1~\lstar\, for all sources, and \macc\, 
of the order $10^{-7}-10^{-9}$ \msunyr, in the range of values commonly obtained for Class II objects. 
The mass accretion rates derived in Cha I are roughly proportional to \mstar$^2$, in agreement with the results found in other low-mass star-forming regions.
We find that the discrepancies observed in the case of \lacc(\brg) and \lacc(\pab) can be related to different intrinsic \pab/\brg\, ratios. The derived ratios 
point to the existence of two different emission modalities, one that agrees with predictions of both wind and accretion models, 
the other suggesting optically thick emission from relatively small regions (10$^{21}$-10$^{22}$ cm$^{-3}$) with gas at low temperatures ($<$4000 K), 
the origin of which needs additional investigation.} 
{}

\keywords{Stars: formation -- Stars: evolution -- Infrared: stars -- Accretion, accretion disks -- Surveys}
\maketitle

\section{Introduction}
\label{sec:intro}

One of the major observational challenges in star-formation studies is the derivation
of the relevant physical parameters characterising the accretion process in young stellar objects (YSOs) 
and the determination of their evolutionary status.

Traditionally, YSOs are divided into evolutionary classes that are empirically based on
the shape of their spectral energy distribution (SED) \citep[Class 0, I, II, and III objects,][]{lada84,andre93}.
This classification reflects the evolution of the source circumstellar material,
from a massive and thick envelope in Class 0 to a tenuous circumstellar disc in Class III,
but it does not give direct information on the accretion activity of the objects, nor on how this is
related to the source/disc system and how it evolves with time. There is indeed evidence
that only a fraction of Class I sources have mass accretion rates high enough to put them
in their main stage of mass-building \citep{enoch09,evans09,antoniucci08,beck07,nisini05a,white04},
while there is a wide difference in disc activity
among Classical T Tauri Stars (CTTSs, Class II objects) of the same mass \citep[e.g.][]{gullbring98}. 
This evidence indicates that the standard
scenario in which most of the object mass is accumulated during the Class 0/I phase
has to be revised: possible alternative pictures are that either the stellar mass is
assembled within the Class 0 stage, or YSOs undergo periods of eruptive events of enhanced mass
accretion. To distinguish the possible interpretations, it is necessary to study large samples of sources of different classes
and to compare their mass and age with their accretion properties. 

From the observational point of view, important quantitative information on the 
active inner and complex region of interaction between the accretion disc, the young star, and the 
jet/wind formation region can be obtained with observations of the many
diagnostic emission/absorption lines available in the optical/near-IR wavelength range.
Several spectroscopic studies have been carried out on samples
of young objects with the aim of deriving their main stellar parameters and
of inferring relevant disc parameters.
In particular, the mass accretion rate through the disc, i.e. the main parameter
regulating the accretion phase of a YSO, has been measured on 
samples of CTTSs with a wide range of masses \citep[e.g.][]{muzerolle98a, natta06, herczeg08, fang09}.
These results were mainly obtained using optical (e.g. \caii\, and \ha\,) and IR (\pab\, and \brg) emission lines to measure the mass accretion rate,
adopting relationships between accretion luminosity and line flux luminosities 
calibrated on direct tracers of accretion like the excess UV emission over the
stellar photosphere \citep[e.g.][]{gullbring98}. More limited quantitative studies of this kind have been performed
in the IR on embedded Class I sources, for which the stellar parameters are much more
difficult to constrain owing to the considerable IR veiling \citep{nisini05b,antoniucci08,connelley10}.

These previous works were mainly based on observations performed
on limited spectral ranges, which sometimes covered only individual diagnostic lines,
which in turn made the results dependent on the choice of the tracer. In addition, 
most studies have been performed on non-homogeneous samples of CTTSs with similar ages, 
which means that no attempt has been performed to compare samples of YSOs of different ages
in star-forming regions with different general characteristics.

In this framework, and to derive quantitative indicators of YSO activity, 
we have undertaken a combined optical/IR unbiased spectroscopic survey on a 
flux-limited sample of Spitzer selected Class I/II sources hosted in six different nearby 
clouds (namely Cha I-II, L1641, Ser, Lup, Vel, and CrA), using the EFOSC2/SOFI instruments mounted on the ESO-NTT telescope. A 
total number of about 150 sources have been observed in a wide wavelength range 
(from 0.6 to 2.4$\mu$m) at low resolution (R$\sim$700) during this project, which has been named
POISSON: \textit{Protostellar Objects IR-optical Spectral Survey On NTT}.
The present article is the first of a series on the analysis of the POISSON data.

In this paper we report on the observations performed on 47 YSOs in the Chamaeleon I 
and II molecular clouds \citep[d=160 pc and 178 pc,][]{whittet97}. 
The Chamaeleon complex consists of three main dark clouds, namely Cha I, Cha II, 
and Cha III \citep{schwartz91}, which present very different star-formation efficiencies (SFE).
In particular, Cha I is the most active of the three clouds,
with an SFE of $\sim$10\% \citep{luhman07,mizuno99}, while the SFE estimated for Cha II is only $\sim$1-3\%.
Observations of the Cha I and II samples will therefore allow us 
to compare properties of sources located at the same distance, but with a 
different star-formation activity.

The population of the Cha I/II clouds has been extensively studied in the past \citep[see
review by][and references therein]{luhman08a}. 
Recent Spitzer observations have been 
combined with near-IR photometry to take a complete census of the YSO population in the two clouds 
\citep[see e.g.][]{young05,allers06,porras07,damjanov07,alcala08,luhman05,luhman08a,luhman08b}.
Optical spectroscopic follow-up has then been used to derive the stellar parameters
of the identified YSOs \citep{luhman07,spezzi08}. 
The estimated mean age of the stellar population is 2 Myr for Cha I \citep{luhman08a} and
3-4 Myr for Cha II \citep{spezzi08}.
Infrared spectroscopy of a few sources of the Cha I/II clouds 
have been analysed by \citet{gomez02} and \citet{natta04}, while 
accretion rates for limited sub-samples have been 
derived using the luminosity of \caii\, and \hi\, IR lines, or the H$\alpha$ 10\% line 
\citep{natta04,mohanty05,spezzi08}. 

Taking advantage of this huge and detailed database of information available for the YSO population
in Cha I/II, we have investigated here he accretion properties of these sources by
using and comparing the existing empirical relationships between line luminosity 
and accretion luminosity. Thanks to the wide range of wavelengths covered by 
our observations, we have been able to measure the accretion luminosity (\lacc) through different tracers 
in a homogeneous way and, more importantly, to compare the \lacc\, determinations 
independently of the source variability. 
We have therefore focused on the analysis of the results obtained with the different lines,
discussing the consistency of the empirical relationships used and the 
reliability of the considered tracers. Part of the analysis has also focused
on the derivation of the excitation conditions pertaining to the IR HI lines,
because these lines are the only tracers of accretion that can be used in both CTTSs and
embedded sources. 

The paper is structured as follows: in Sects.~\ref{sec:sample} and \ref{sec:parameters} we introduce the 
sample and present the main source-properties derived from the literature. In Sect.~\ref{sec:observations} 
we describe the observations and data reduction. In Sect.~\ref{sec:spectra} we present the spectra 
and the detected features. Then, we derive the accretion luminosity of the 
objects from the different tracers in Sect.~\ref{sec:accretion}, comparing and discussing the obtained values. The analysis of 
the observed \hi\, \pab/\brg\, ratio, which provides information on the \hi\, emission mechanism, is given 
in Sect.~\ref{sec:ratios}. The main conclusions of our work are finally summarised in Sect.~\ref{sec:conclusions}.

\section{Sample selection}
\label{sec:sample}

The sample has been selected from the young population of ClassI/II objects
identified through Spitzer surveys of the Chamaeleon I and II molecular
clouds. Reference works were \citet{luhman07} and \citet{luhman08a} for Cha I,
and \citet{alcala08} and \citet{spezzi08} for Cha II. 
From these surveys we selected the sources with an SED spectral index between 2 and 24 \um\, $\alpha_{2-24} \gtrsim -1$
and a $K$-band magnitude $<$12 mag, constrained by instrumental sensitivity.
This selection resulted in a total of 29 (15) targets for the Cha I (Cha II) cloud, namely 2 (1) Class I, 4 (0) flat and 23 (14) Class II
objects.
Three more Class IIs with $\alpha < -1$ (1 in Cha I and 2 in Cha II) were found to be very close to some of the selected targets and were accordingly observed 
with the same slit acquisition. These sources were eventually included in the sample, which therefore amounts to a total of 47 objects.

The very small number of Class I sources is due to the relatively old age of the population of these clouds, with respect to other 
nearby molecular clouds such as $\rho$ Oph.

The complete list of targets is given in Tables \ref{tab:chai_prop} and \ref{tab:chaii_prop} together with their
main parameters, which are presented in the following section.

\section{Source parameters}
\label{sec:parameters}

\subsection{Spectral type, mass, and luminosity.}
\label{sec:st}
Spectral classification of most of the objects has been performed on the basis of optical spectroscopy by \citet{luhman04,luhman07} for Cha I and by \citet{spezzi08} for Cha II.
The spectral types, stellar masses, and luminosities derived by these authors are reported in Tables \ref{tab:chai_prop} and \ref{tab:chaii_prop}.
Basically, all sources present late spectral types (K, M) and masses ranging between 0.2 and 1.4 M$_{\odot}$, with the exception of the two 2M$_{\odot}$ sources DI Cha 
(Cha I-8, spectral type G2) and DK Cha (Cha II-1, spectral type F0).

For each source we also computed the total luminosity between 1 and 70 or 100 \um\, (depending on the longest wavelength  photometric measurement available), which
we indicate as $L_{IR}$, by integrating the SED\footnote{The computation was performed starting from the $J$-band value and considering straight lines 
(in the Log ($\lambda$)-Log ($\lambda F_\lambda$) plan) between available SED points corrected for extinction.} 
constructed with photometry from the 2MASS survey \citep{2mass} and Spitzer instruments IRAC and MIPS \citep[see][]{alcala08,luhman08a}.
We notice that for almost all sources $L_{IR}$ is very similar or comparable (within a factor of 2) with $L_*$.
We may interpret this finding as another indication that most of the objects are quite evolved sources characterised by weak infrared excesses and 
presenting a luminosity that is basically dominated by the stellar photospheric emission. We note that the few sources where \lstar\, and $L_{IR}$ are significantly 
different display either relatively early spectral types (e.g. Cha I-8) or high extinctions (e.g. Cha II-1, Cha II-3).

The spectral classification, and consequently the relative stellar parameters, were not available for seven objects of the Cha I cloud. 
Noticeably, these are the reddest sources of our Cha I sample (i.e. displaying the highest values of the spectral index $\alpha_{2-24\mu m}$) 
and are probably more embedded than the others (see Sect.~\ref{sec:extinction}). 
This is consistent with the difficulties \citet{luhman04} had to obtain a good quality optical spectrum and derive a spectral type for these objects.
In the remainder of the paper we will refer to our estimate of $L_{IR}$ in place of the unknown $L_*$ for these seven sources. However,
because these are the objects of the sample characterised by the strongest IR excess, $L_{IR}$ is likely to overestimate
$L_*$ in these cases.

\subsection{Extinction}
\label{sec:extinction}
The knowledge of the amount of extinction toward the sources is fundamental to correctly de-redden line fluxes, 
which will eventually be used to derive the accretion luminosities (see Sect.~\ref{sec:lacc}). 
Extinction estimates for our targets are available in the literature for both Cha I
\citep[$A_J$ values from][]{luhman04} and Cha II \citep[$A_V$ values from][]{spezzi08}.

Investigations of the interstellar medium in the Chamaeleon regions have shown that the value of the total-to-selective extinction ratio $R_V$ 
is fairly high in both clouds: greater than 5 in Cha I \citep{luhman04} and up to 5-6 in the densest parts of Cha II \citep[e.g][]{vrba84,whittet94,whittet97}. 
On this basis, we have adopted the extinction law of \citet{cardelli89} with $R_V$=5.5 to transform the provided $A_J$ and $A_V$ values 
to obtain the extinction at the wavelength of interest.
If we had used the standard extinction law of \citet{rieke85} to derive for example the extinction in the $K$-band from the provided $A_V$, we would have obtained  
a variation of about 20\% for $A_K$.
On average, the targets display relatively low extinction values ($A_V < 4$ for 14 objects in Cha I and 15 in Cha II, see Tables~\ref{tab:chai_prop} and \ref{tab:chaii_prop}), which generally agrees with the negative spectral index $\alpha_{2-24\mu m}$ measured in most of the sources.
 
No extinction value was given by Luhman for the seven Cha I objects without spectral classification, four of which present \brg\, line emission.
For these sources we therefore derived a rough estimate of $A_V$
from their position on the 2MASS $J-H$ vs $H-K$ colour-colour diagram. Adopting the extinction law of \citet{cardelli89} with $R_V$=5.5, 
we de-reddened the objects to bring them back on the CTTS locus \citep{meyer97}. We find quite high extinctions ($A_V>6$) for all sources except
for Cha I-15, which conversely presents near-IR colours compatible with an extinction close to zero.

\begin{sidewaystable*}[h!]
\begin{tiny}
\caption{Main properties of Cha I targets. Stellar parameters L$_{*}$, R$_{*}$, M$_{*}$, ST and extinction are taken from \citet{luhman04,luhman07}, unless explicitly noted.}
\label{tab:chai_prop}
\begin{tabular}{l|lcclcccccccc}
\hline
\hline
Id & Name & $\alpha^{a}$ (2000) & $\delta^{a}$ (2000) & Other Names & $\alpha_{2-24\mu m}$ & L$_{IR}^{b}$ & L$_{*}$ & R$_{*}$ & M$_{*}$ & ST &  A$_{V}^{c}$ & m$_K^{d}$ \\ 
   &       &                  &         &         &   &   L$_{\sun}$   &L$_{\sun}$&R$_{\sun}$ & M$_{\sun}$&    & mag     &   mag \\
\hline
Cha I-1&    T11&                              11:02:24.91&    -77:33:35.7&   Sz9, CS Cha&               -0.72&    0.96&        1.2&         2.07&        0.87&                           K6&                     0.21&        8.20\\                   
Cha I-2&    CHSM1715&                         11:04:04.25&    -76:39:32.8&   CHSM1715&                  -0.62&    0.08&        0.05&        0.71&        0.2&                            M4.25&                  2.22&        10.90\\                   
Cha I-3&    T14&                              11:04:09.09&    -76:27:19.3&   Sz11, CT Cha&              -0.51&    0.68&        0.95&        1.72&        1.07&                           K5&                     1.35&        8.66\\                   
Cha I-4&    ISO52&                            11:04:42.58&    -77:41:57.7&   ISO52, B18&                -0.84&    0.07&        0.09&        0.94&        0.25&                           M4&                     1.08&        10.64\\                   
Cha I-5&    Hn5&                              11:06:41.80&    -76:35:48.9&   Hn5&                       -0.82&    0.13&        0.11&        1.08&        0.22&                           M4.5&                   0.96&        10.13\\                   
Cha I-6&    ISO92&                            11:07:09.19&    -77:23:04.9&   IRAS11057-7706&         	0.71&    0.60&        -&           -&           -&                              -&                      6.0$^f$&           10.92\\                   
Cha I-7&    ISO97&                            11:07:16.22&    -77:23:06.8&   -&                      	0.15&    0.09&        -&           -&           -&                              -&                      10.0$^f$&           11.68\\                   
Cha I-8&    T26&                              11:07:20.74&    -77:38:07.3&   Sz19, DI Cha&              -0.92&    4.00&        12&          3.37&        2&                              G2&                     2.25&        6.22\\                   
Cha I-9&    B35&                              11:07:21.42&    -77:22:11.7&   ISO101&              	   -0.05&    0.18&        -&           -&           -&                         -&                      20.0$^f$&           10.93\\                   
Cha I-10&   CHXR30B&                          11:07:57.30&    -77:17:26.2&   B38&                       -0.79&    0.24&        0.22&        1.16&        0.58&                           M1.25&                  9.47&        9.95\\                   
Cha I-11&   T30&                              11:07:58.09&    -77:42:41.3&   Sz23&                      -0.91&    0.16&        0.15&        1.06&        0.44&                           M2.5&                   3.77&        9.89\\                   
Cha I-12&   CHXR30A&                          11:08:00.02&    -77:17:30.4&   CHXR30A&                   -1.63&    0.32&        1.4&         2.52&        0.95&                              K8&                     8.93&        9.09\\                   
Cha I-13&   T31&                              11:08:01.48&    -77:42:28.8&   Sz24, VW Cha&              -0.98&    1.93&        3&           3.7&         0.74&                           K8&                     2.16&        6.96\\                   
Cha I-14&   Cha\_IRN&                         11:08:38.96&    -77:43:51.3&   C9-2,ISO150	&   			0.84$^{e}$&    2.02&        -&           -&           -&                    $<$M0&                  11.0$^f$&           8.71\\                   
Cha I-15&   C9-3&                             11:08:42.96&    -77:43:50.0&   -&                         -0.79&    0.12&        -&           -&           0.25&                           -&                      $\sim0^f$&           11.78\\                   
Cha I-16&   T38&                              11:08:54.64&    -77:02:12.9&   Sz29, VY Cha&              -0.88&    0.23&        0.34&        1.36&        0.66&                           M0.5&                   2.70&         9.46\\                   
Cha I-17&   CHXR79&                           11:09:18.12&    -76:30:29.2&   ISO186&                    -0.91&    0.37&        0.55&        1.84&        0.58&                           M1.25&                  5.75&        9.07\\                   
Cha I-18&   C1-6&                             11:09:22.66&    -76:34:32.0&   CED112-IRS2&               -0.49&    0.90&        0.8&         2.21&        0.58&                           M1.25&                  9.79&        8.67\\                   
Cha I-19&   C1-25&                            11:09:41.92&    -76:34:58.4&   ISO199&          		   -0.23&    0.30&        -&           -&           -&                              -&                      14.0$^f$&           10.00\\                   
Cha I-20&   Hn10-e&                           11:09:46.21&    -76:34:46.3&   FL2004 55&                 -0.46&    0.15&        0.15&        1.13&        0.34&                           M3.25&                  3.02&        10.05\\                   
Cha I-21&   B43&                              11:09:47.42&    -77:26:29.0&   ISO207&                    -0.91&    0.15&        0.22&        1.37&        0.36&                           M3.25&                  6.77&        10.24\\                   
Cha I-22&   T42&                              11:09:53.40&    -76:34:25.5&   ISO223,Sz32,FM Cha &   	   -0.55$^{e}$&    3.61&        3&           3.05&        0.75&                              K5&                     4.40&        6.46\\                   
Cha I-23&   T43&                              11:09:54.07&    -76:29:25.3&   Sz33, CHXX12&              -0.87&    0.29&        0.48&        1.82&        0.54&                           M2&                     4.40&        9.25\\                   
Cha I-24&   C1-2&                             11:09:55.05&    -76:32:40.9&   ISO226, CED112-IRS5 &       0.03&     0.57&        -&           -&           -&                              -&                      14.0$^f$&           9.67\\                   
Cha I-25&   Hn11&                             11:10:03.69&    -76:34:58.4&   ISO232&                    -0.64&    0.36&        0.66&        1.73&        0.76&                           K8&                     6.41&        9.44\\                 
Cha I-26&   ISO237&                           11:10:11.41&    -76:35:29.2&   ISO237&                    -0.58&    0.59&        1.2&         2&           0.95&                           K5.5&                   5.75&        8.62\\                   
Cha I-27&   T47&                              11:10:49.59&    -77:17:51.7&   Sz37, HBC584&              -0.57&    0.31&        0.42&        1.71&        0.53&                           M2&                     3.50&        9.18\\                   
Cha I-28&   ISO256&                           11:10:53.59&    -77:25:00.4&   -&                         -0.59&    0.07&        0.07&        0.86&        0.21&                           M4.5&                   7.64&        11.34\\                   
Cha I-29&   T49&                              11:11:39.65&    -76:20:15.2&   Sz39, XX Cha&              -0.92&    0.31&        0.37&        1.6&         0.53&                           M2.5&                   1.02&        8.87\\                   
Cha I-30&   T53&                              11:12:30.92&    -76:44.24.1&   Sz43, CW Cha&              -1.05$^{e}$&    0.33&        0.39&        1.52&        0.5&                         M1&                     2.37&        9.13\\                   
\hline
\end{tabular} 
~\\
~\\
Notes. \\
$^{a}$: coordinates from the 2MASS catalogue. \\
$^{b}$: luminosity computed by integrating the spectral energy distribution between 1 and 70/100\um, using 2MASS and Spitzer measurements corrected for extinction.\\
$^{c}$: obtained from A$_J$ given by \citet{luhman07}. \\
$^{d}$: $K$-band magnitude from 2MASS.\\
$^{e}$: spectral index computed from 2 to 12 \um\, \citep[see][]{luhman08a}.\\
$^{f}$: extinction estimate from the $(J-H)$-$(H-K)$ colour-colour diagram.
\end{tiny}
\end{sidewaystable*}

%
\begin{sidewaystable*}[h!]
\begin{tiny}
\caption{Main properties of Cha II targets. Stellar parameters L$_{*}$, R$_{*}$, M$_{*}$, ST and extinction are taken from \citet{spezzi08}, unless explicitly noted.}
\label{tab:chaii_prop}
\begin{tabular}{l|lcclcccccccc}
\hline
\hline
Id & Name & $\alpha^{a}$ (2000) & $\delta^{a}$ (2000) & Other Names & $\alpha_{2-24\mu m}$ & L$_{IR}^{b}$ & L$_{*}$ & R$_{*}$ & M$_{*}$ & ST & A$_{V}$ & m$_K^{c}$ \\ 
   &       &                  &         &         &   &   L$_{\sun}$   &L$_{\sun}$&R$_{\sun}$ & M$_{\sun}$&    & mag      &    mag      \\
\hline
Cha II-1&   DK Cha&                           12:53:17.22&    -77:07:10.6&   IRAS12496-7650,HH274&             -0.72&    53.7&       18.62&       2.77&        2$^{d}$&                             F0&          10.55&                  5.19\\                   
Cha II-2&   IRAS 12535-7623&                  12:57:11.72&    -76:40:11.1&   CHIIXR2&                          -0.98&    1.16&        1.38&        2.71&        1.05&                           M0&          3.36&                   8.40\\                   
Cha II-3&   ISO28&                            12:59:06.56&    -77:07:40.1&   IRAS12553-7651 &  				   0.56&    2.29&        15.85&       5.58&        0.7$^{d}$&                           K4.5&        38.9&                   10.66\\                   
Cha II-4&   Sz49&                             13:00:53.23&    -76:54:15.1&   ISO55&                     		  -0.61&    0.19&        0.2&         1.03&        0.75&                           M0.5&        2.28&                   10.63\\                   
Cha II-5&   Sz48SW&                           13:00:53.46&    -77:09:08.6&   CHIIXR7&                          -0.95&    0.09&        0.26&        1.25&        0.7&                            M1&          3.87&                   9.45\\                   
Cha II-6&   Sz50&                             13:00:55.32&    -77:10:22.2&   ISO52,CHIIXR8&              	  -0.88&    0.80&        1.15&        3.1&         0.5&                            M3&          3.78&                   8.85\\                   
Cha II-7&   CM Cha&                           13:02:13.51&    -76:37:57.7&   CHIIXR13,IRAS 12584-7621&         -0.95&    0.83&        0.72&        1.78&        1.15&                           K7&          1.52&                   8.52\\                   
Cha II-8&  IRAS13005-7633&                   13:04:22.84&    -76:50:05.5&   Hn22&                             -0.66&    0.31&        0.23&        1.24&        0.5&                            M2&          0.61&                   9.73\\                   
Cha II-9&  Hn23&                             13:04:24.10&    -76:50:01.2&   -&                                -1.03&    0.87&        0.87&        1.6&         1.4&                            K5&          1.24&                   8.77\\                   
Cha II-10&  Hn24&                             13:04:55.71&    -77:39:49.5&   -&                                -0.97&    0.76&        1.05&        2.37&        1&                              M0&          2.76&                   8.92\\                   
Cha II-11&  Sz53&                             13:05:12.69&    -77:30:52.5&   -&                                -1.07&    0.28&        0.32&        1.39&        0.75&                           M1&          3.68&                   9.93\\                   
Cha II-12&  Sz56&                             13.06:38.82&    -77:30:35.2&   -&                                -1.08&    0.20&        0.34&        1.78&        0.3&                            M4&          3.18&                   10.41\\                   
Cha II-13&  Sz57&                             13:06:56.56&    -77:23:05.4&   CHIIXR34&                         -1.55&    0.28&        0.41&        2.21&        0.15&                           M5&          3.09&                   9.80\\                   
Cha II-14&  Sz58&                             13:06:57.44&    -77:23:41.5&   IRAS 13030-7707&                  -0.93&    0.81&        0.69&        1.43&        1.2&                            K5&          3.87&                   8.76\\                   
Cha II-15&  Sz61&                             13:08:06.28&    -77:55:05.1&   BM Cha&                           -0.97&    1.52&        1.17&        1.87&        1.4&                            K5&          3.13&                   7.95\\                   
Cha II-16&  IRASF 13052-7653NW&               13:09:09.87&    -77:09:43.7&   -&                                -1.07&    0.31&        0.2&         1.03&        0.75&                           M0.5&        2.28&                   9.64\\                   
Cha II-17&  IRASF 13052-7653N &               13:09:10.71&    -77:09:44.3&   -&                                -1.17&    0.20&        0.34&        1.49&        0.62&                           M1.5&        0.41&                   9.07\\                   
\hline
\end{tabular} 
~\\
~\\
Notes.\\
$^{a}$: coordinates from the 2MASS catalogue. \\
$^{b}$: luminosity computed by integrating the Spectral Energy Distribution between 1 and 70-100\um, using 2MASS and Spitzer measurements corrected for extinction.\\
$^{c}$: $K$-band magnitude from 2MASS.\\
$^{d}$: mass estimate from \citet{dantona97} evolutionary tracks \citep[see][]{spezzi08}.
\end{tiny}
\end{sidewaystable*}

\section{Observations and data reduction}
\label{sec:observations}

Our observations were carried out at the ESO New Technology Telescope \citep{ntt} with the EFOSC2 and SofI
spectrographs \citep{sofi,efosc2} on 10-14 February 2009.
We used the EFOSC2 grism \#16 equipped with the 0$\farcs$7 slit (R $\sim$ 700), and the red and blue grisms 
(hereafter RG and BG) of SofI in combination with the 0$\farcs$6 slit (R $\sim$ 900). 
The three low-resolution spectra span the 0.60-1.02, 0.94-1.65, and 1.50-2.40 \um\, intervals, 
respectively, and accordingly provide a complete coverage of the 0.6-2.4 \um\, wavelength range.

We acquired the RG spectrum for all sources of the sample, while the BG and EFOSC2 (hereafter ``optical") spectra were taken only for 
the brightest objects with $J <$ 14 and $I < $ 15, respectively. 
Hence, the final dataset consists of 47 RG, 38 BG, and 33 optical spectra (see Table~\ref{tab:chai_lines} and \ref{tab:chaii_lines}).
All data were reduced using the IRAF \footnote{IRAF (Image Reduction and Analysis Facility) is a general-purpose software system 
for the reduction and analysis of astronomical data. 
IRAF is written and supported by the IRAF programming group at the National Optical Astronomy Observatories (NOAO) in Tucson, 
Arizona. \textit{http://iraf.noao.edu}} software package. 

For the infrared spectra we followed the standard procedures for bad pixel removal, flat-fielding, and sky-subtraction. 
Spectra of telluric standard stars were acquired at air-masses similar to those of the targets
and were used after removal of any intrinsic line to correct the scientific spectra for atmospheric absorption. 
Wavelength calibration for all spectra was obtained using Xenon-Argon arc lamps. 

Strong seeing variations and instrumental problems (the source did not remain correctly centred on the slit passing from the A to B positions during the nodding) 
caused flux losses that prevented us from calibrating the flux scale of the spectra using spectro-photometric standards. 

Hence, we adopted the 2MASS $K$-band magnitude for the calibration of the RG spectral segment. Then, we normalised the BG segment 
to obtain a good match of the two spectra in the spectral region where they overlap.
Once this procedure was applied, we checked the consistency of our calibrated spectra with the 2MASS $J$ and $H$ points and found  
a very good agreement for most of the objects ($\Delta (H-K)$ and $\Delta (J-K) \leq 0.1$ mag), with maximum colour discrepancies 
of $\Delta (H-K) = 0.3$ and $\Delta (J-K) = 0.4$ mag.
This is an indication that there have been no strong variations of the infrared colours of the sources since the 2MASS measurements.

The EFOSC2 spectra were reduced performing the standard procedures for dark current subtraction, bad pixel and 
fringing removal, and flat-fielding. The wavelength calibration was carried out using helium-argon arc lamps. 
The spectra acquired with grism \#16 display an evident drop owing to an efficiency loss above about 0.9 \um . 
This effect could not be properly calibrated out using spectro-photometric standard
star spectra because it appears to be strongly dependent on the colour of the observed source, so that it turns out to be very different for the standards with respect to our scientific targets, which are much redder. We therefore decided to cut the optical spectra at 0.9 \um. 
Because of flux losses caused by seeing variation during the acquisitions, we preferred to use the BG spectrum as a reference to flux-calibrate the optical segment.
The agreement we find with available $I$ band photometry is good for the vast majority of the sources and the maximum colour variation we observe
is $\Delta (I-K) = 0.5$ mag.

Although the contiguity and partial overlapping of the spectral segments ensures a good inter-calibration of the spectra, the absolute calibration of the 
flux scale suffers from the 
uncertainty of the value of the $K$-band magnitude at the moment of observations with respect to the 2MASS $K$-value, which was eventually used as reference. 
Observations show that young stellar objects display a $K$-band mean variability that can be estimated to be of the order of 0.5 mag over 
time scales of years \citep[see e.g.][]{alves_de_oliveira08}. 
Consequently, we will assume this value, which corresponds to a factor of about 1.6 in flux, as an estimate of the error for 
the flux calibration of our spectra.

\begin{figure*}
\centering
\includegraphics[width=9.0cm]{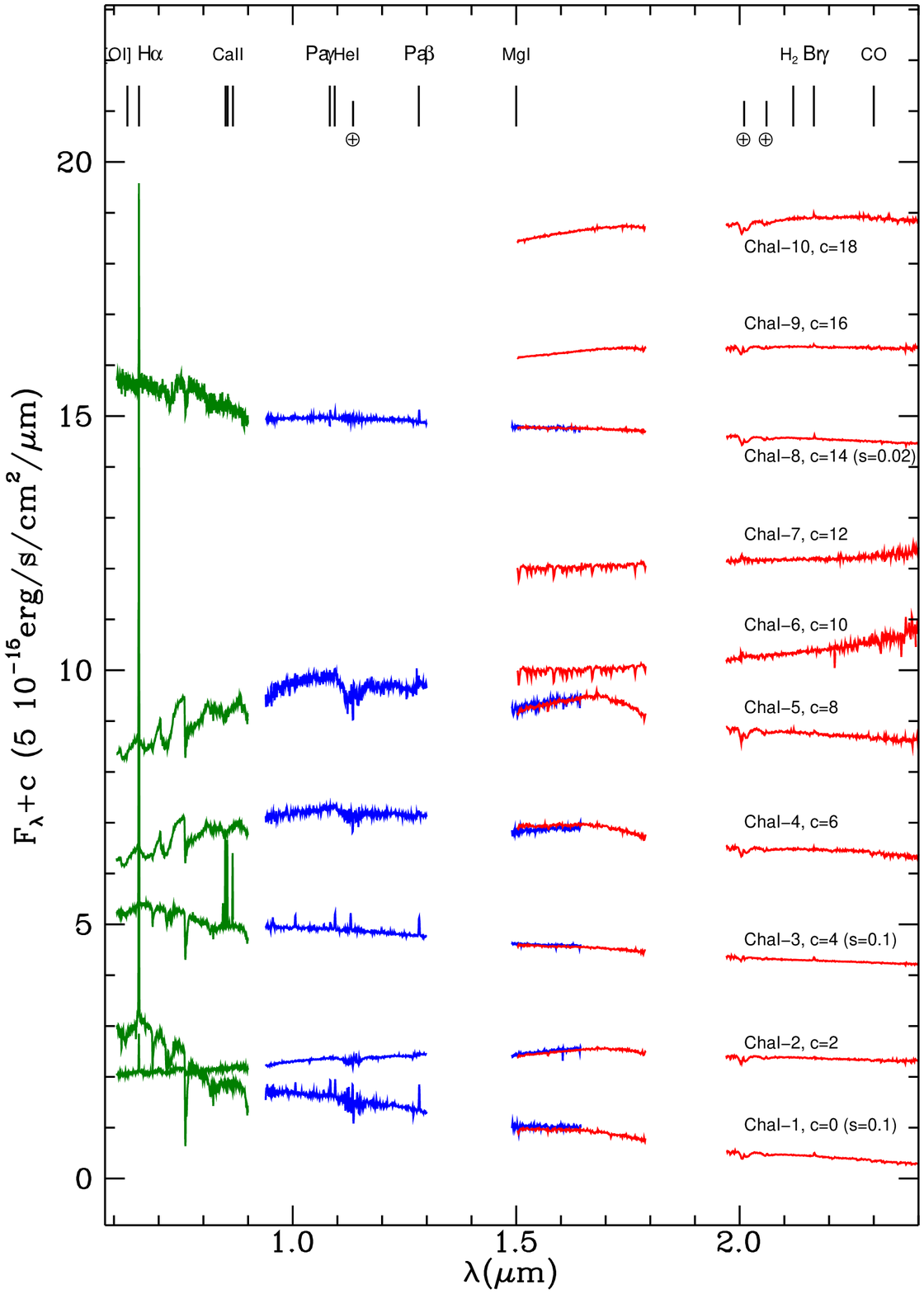}
\includegraphics[width=9.0cm]{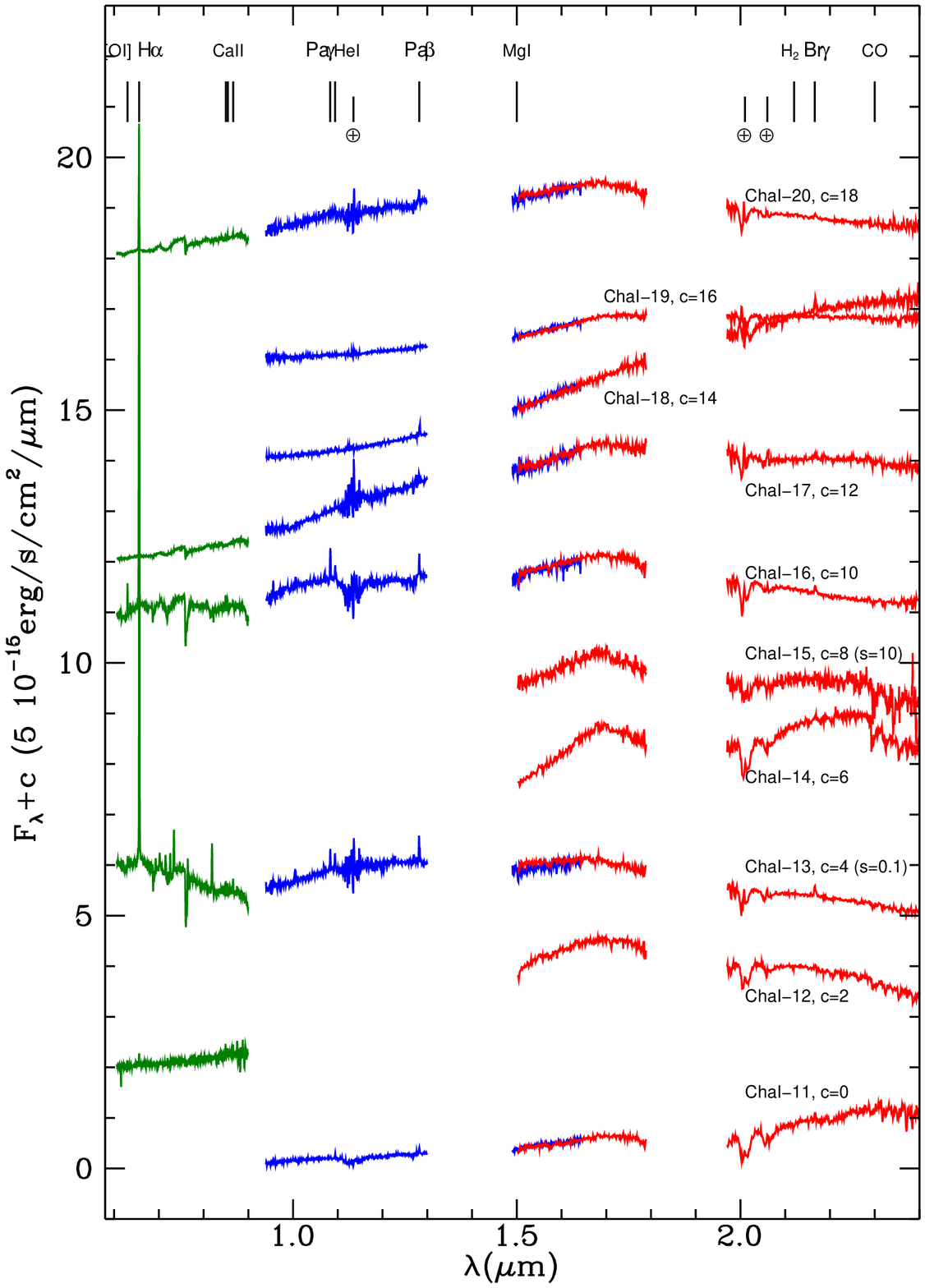}\\[-2ex]
\includegraphics[width=9.0cm]{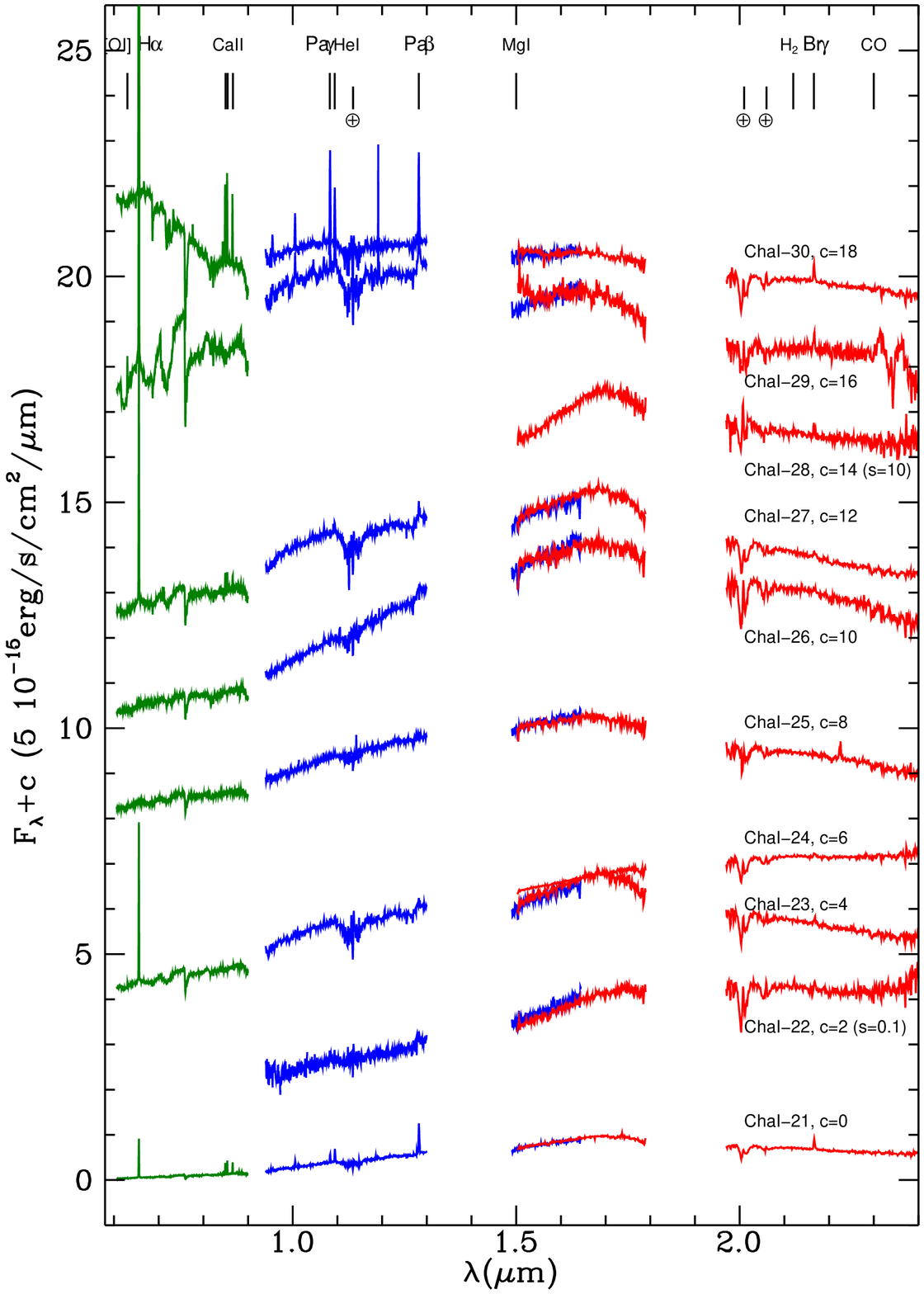}
\caption{\label{fig:spectra1} Spectra of the Cha I sources. The three parts of the spectrum are displayed in different colours 
(EFOSC2 grism \#16 in green, SofI blue grism in blue, and SofI red grism in red). The spectra were offset 
and some of them were multiplied by a scale factor $s$ (indicated in parentheses) for better visualisation.  
Wavelength intervals heavily corrupted by atmospheric absorption were removed. The position of the main emission lines 
present in the covered spectral range is indicated.}
\end{figure*}
\begin{figure*}

\centering
\includegraphics[width=9.0cm]{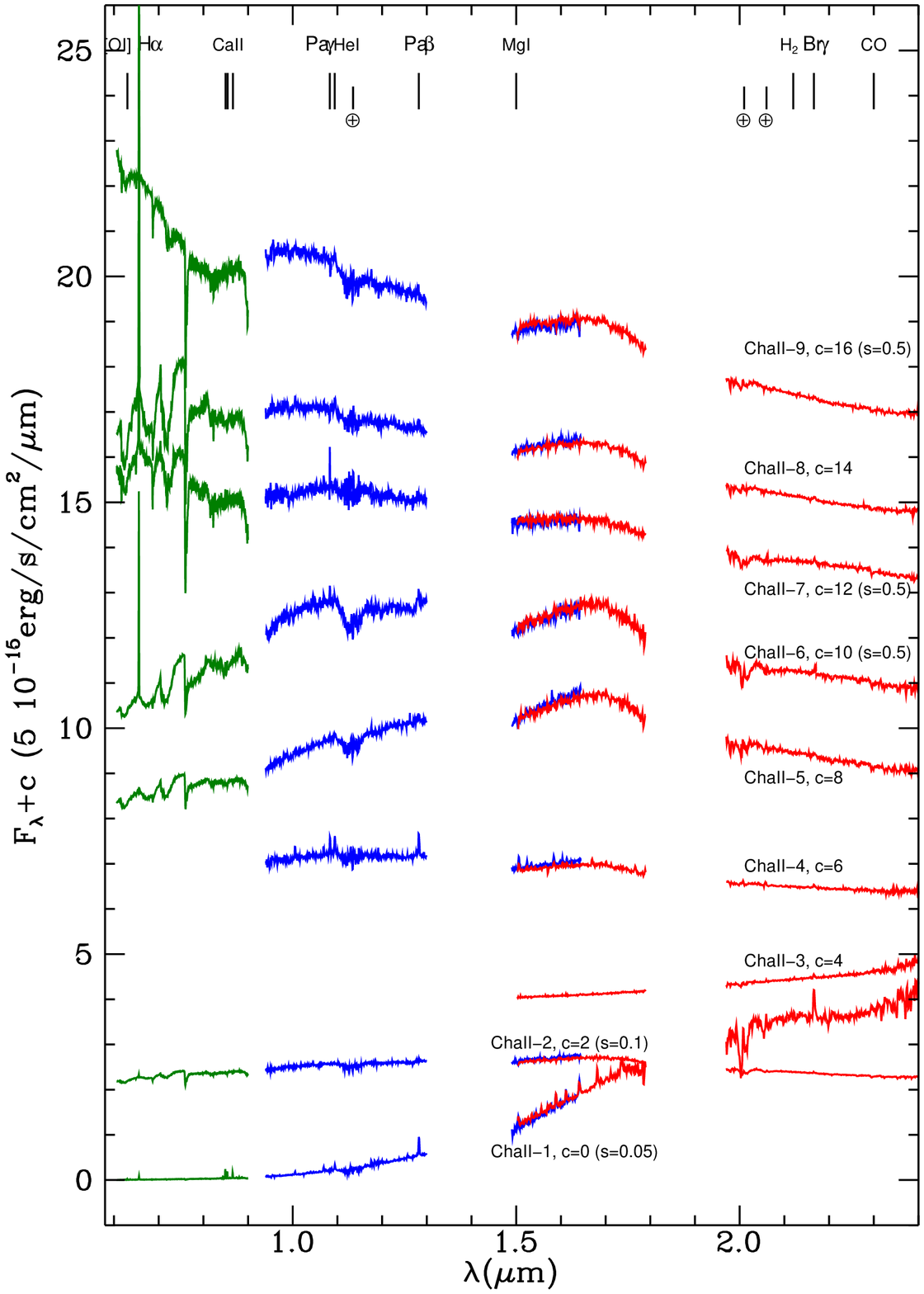}
\includegraphics[width=9.0cm]{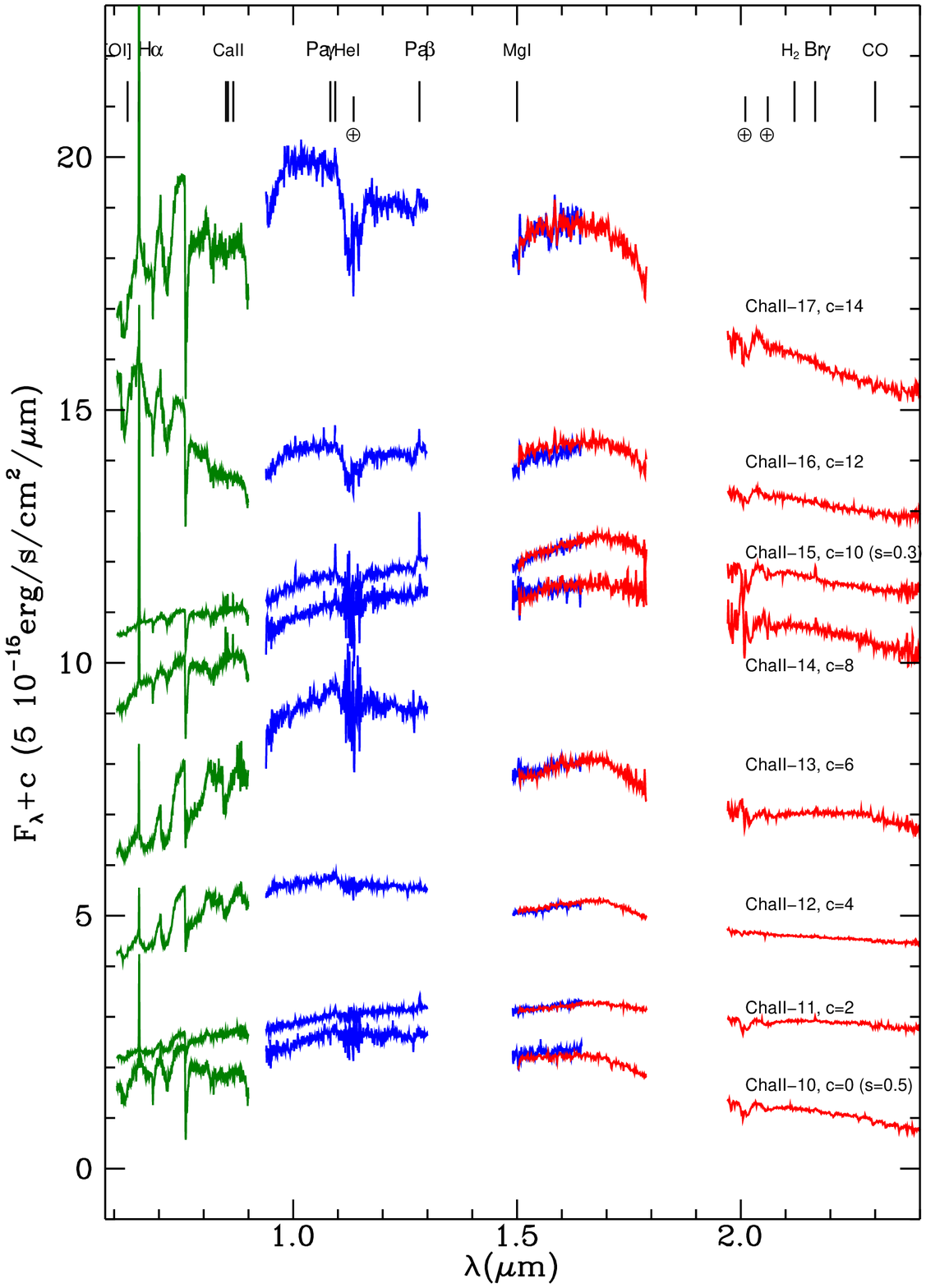}
\caption{ \label{fig:spectra2} Spectra of the Cha II sources (see caption of Fig.~\ref{fig:spectra1}).} 
\end{figure*}

\section{Spectra}
\label{sec:spectra}

The spectra of the sources of the two samples are displayed in Figs.~\ref{fig:spectra1} and \ref{fig:spectra2}.
They show several emission lines that remain unresolved at the low spectral resolution of the observations. 

We detect many permitted emission lines commonly found in young sources: the most numerous features are the \hi\, 
recombination lines from the Paschen and Brackett series (in particular 
\pab\, and \brg) and \ha\, at optical wavelengths. \brg\, is detected in 39 sources out of the 47 composing the sample.
Other permitted lines we observe are the \hei\, line at 1.08 \um, which is commonly associated with emission in 
expanding stellar winds \citep[e.g.][]{edwards06,ray07}, the \caii\, triplet (8498, 8542, 8662 \AA), and the \ion{Mg}{i} at 1.5 \um. 

ChaII-1 (DK Cha) displays the richest emission spectrum in the sample: this is also the 
only object showing significant emission from \nai\, (2.2\um) and ro-vibrational CO transitions (from 2.3\um\, on).
Both features are usually associated with the presence of active circumstellar discs \citep[e.g.][]{najita03,antoniucci08}.
This interesting object is analysed in detail in a dedicated paper \citep{garcia_lopez11}.

The forbidden emission lines of \oi, \sii, \feii\, are classical indicators of outflow activity in young stellar 
objects \citep[e.g.][]{reipurth01,nisini05b,podio06}. While the \oi\, line at 6300 \AA\, is detected in 15 objects, the \sii\, emission (6716+6731 \AA) is observed only
in three sources and \feii\, lines (e.g. 1.25 \um, 1.64 \um) are detected only in ChaII-1.

The relative lack of jet-line detections (see Tables~\ref{tab:chai_lines} and \ref{tab:chaii_lines}) can mainly be ascribed to the low resolution and limited sensitivity 
of the observations. This is also consistent with the fact that all these lines are indeed observed in the brightest object (Cha II-1).
For example, on the basis of the \oi/\feii\, ratio observed in various jets, which is in the range 0.1-1 \citep[e.g.][]{nisini05b,podio06},
we can expect intrinsic fluxes of the order 10$^{-14}$-10$^{-15}$ erg s$^{-1}$ cm$^{-2}$ for the \feii\, 
1.25\um\, line, which will be even lower in the observed spectra owing to the extinction. 
Now, considering the typical rms we measure in the $J$ band, we find a (3$\sigma$) 
upper limit of about 2 10$^{-14}$ erg s$^{-1}$ cm$^{-2}$ , which means that our observations are often not 
sensitive enough to detect this line.

High spatial resolution observations for \sii show that the \oi/\sii\, ratio observed very close to the source in many jet-driving YSOs 
is usually greater than a factor 3-5 \citep[e.g.][]{hirth97,melnikov09}, so that where \oi\, is detected at the 10$^{-15}$ erg s$^{-1}$ cm$^{-2}$ level
we expect \sii\, fluxes of order 10$^{-16}$ erg s$^{-1}$ cm$^{-2}$, which are again too low for the sensitivity of our survey.
In agreement with this picture, we remark that \sii\, emission is observed only in sources displaying an \oi\, flux of the order 10$^{-14}$ erg s$^{-1}$ cm$^{-2}$.

The molecular transitions of \htwo\, (in particular the 2.12 \um\, line), which is another typical tracer of shocks from protostellar jets, 
are detected at a significant level only in four sources.
\citet{beck08} found extended \htwo\, emission in several bright CTTSs, but the observed lines are characterised by a low contrast against the continuum,
even at a spectral resolution of about 5000. Hence, in our sample we are probably not sensitive enough to detect \htwo\, lines, except for the
strongest (and/or extended) emission cases. On the other hand, the paucity of \htwo\, features might also suggest a low abundance of molecular gas in the 
close circumstellar environment of many objects.

In Tables \ref{tab:chai_lines} and \ref{tab:chaii_lines} we report for each source the flux of the five emission 
lines that will be used in our analysis to derive the accretion luminosity of 
the objects (namely \ha, \pab, \brg, \oi, and \caii, see Sect. \ref{sec:accretion}), providing upper limits for non-detections, and indicating also the presence 
of other emission features detected at the 3$\sigma$ level.

In general, no evident absorption features are detected in the infrared part of the spectra, as might be 
expected given both the presence of veiling by excess emission, 
typically observed in young sources \citep[e.g.][]{greene96a, nisini05a}, 
and the low resolution of the acquisitions.
Conversely, several objects present broad absorption bands (TiO) at optical wavelengths, which are compatible with
late spectral type (late-K and M) photospheres and agree with the classification available 
in the literature (see Sect. \ref{sec:parameters}).

\begin{sidewaystable*}
\begin{tiny}
\caption{Cha I sample: measured flux (not corrected for extinction) of the five emission lines used as accretion tracers. Equivalent widths are given as useful reference. 
The observed spectral segments are indicated in the second column. Upper limits are provided for non-detections (3$\sigma$ level), whereas the ``...'' indicates 
that the relative spectral segment is not available. Other emission features detected are also indicated. The last column shows the value of the extinction-corrected 
Pa$\beta$/Br$\gamma$ ratio $R$.}
\label{tab:chai_lines}
\begin{tabular}{l|c|cc|cc|cc|cc|cc|c|c}
\hline
\hline
Id & Spectral & \multicolumn{2}{c|}{[\ion{O}{i}] 0.630$\mu$m}  & \multicolumn{2}{c|}{H$\alpha$}  & \multicolumn{2}{c|}{\ion{Ca}{ii} 0.854$\mu$m}  & \multicolumn{2}{c|}{Pa$\beta$}  & \multicolumn{2}{c|}{Br$\gamma$}  & other lines$^{b}$ & $R$\\ 
   & segment$^{a}$ &(F $\pm$ $\Delta$ F)10$^{-14}$ & EW & (F $\pm$ $\Delta$ F)10$^{-14}$ & EW & (F $\pm$ $\Delta$ F)10$^{-14}$ & EW & (F $\pm$ $\Delta$ F)10$^{-14}$ & EW & (F $\pm$ $\Delta$ F)10$^{-14}$ & EW & &\\
   & &erg s$^{-1}$ cm$^{-2}$ & \AA & erg s$^{-1}$ cm$^{-2}$ & \AA & erg s$^{-1}$ cm$^{-2}$ & \AA & erg s$^{-1}$ cm$^{-2}$ & \AA & erg s$^{-1}$ cm$^{-2}$ & \AA & & \\
\hline

 Cha I-1& OBR 	   &                  $<$5.7  &      & 903.0$\pm$        4.8 &  -57.9 &          $<$12.4  &      & 44.3$\pm$        1.7 &   -6.6 & 12.7$\pm$        0.9 &   -5.8 & \ion{He}{I} & 3.6$\pm$0.3\\
 Cha I-2& OBR 	   &        0.25$\pm$        0.07 &   -6.4 &   5.8$\pm$        0.1 &  -81.9 &0.5$\pm$        0.1 &   -4.2 &            $<$0.38  &      &  0.4$\pm$        0.1 &   -2.1 & \ion{He}{I} & $<$1.4 \\
 Cha I-3& OBR 	   &                  $<$3.0  &      & 393.0$\pm$        1.4 &  -56.4 &2.0$\pm$        0.7 &  -23.4 & 42.9$\pm$        1.2 &  -11.1 & 11.8$\pm$        0.6 &   -8.2 & \ion{He}{I} & 4.6$\pm$0.3\\
 Cha I-4& OBR 	   &                  $<$0.20  &      &   1.8$\pm$        0.2 &   -7.7 &          $<$0.79  &      &  2.2$\pm$        0.8 &   -3.9 &  0.3$\pm$        0.1 &      & 
 & 8.8$\pm$4.2\\
 Cha I-5& OBR 	   &                  $<$0.24  &      &  14.7$\pm$        0.1 &  -50.2 &          $<$0.71  &      &  7.2$\pm$        0.1 &   -1.0 &  1.8$\pm$        0.5 &   -4.8 & H$_2$ 
 & 4.8$\pm$1.2\\
 Cha I-6& --\,--\,R  &                  ...  &    ...  &            ...  &    ...  &          ...  &    ...  &            ...  &    ...  &            $<$1.6  &      & 
 & ...\\
 Cha I-7& --\,--\,R  &                  ...  &    ...  &            ...  &    ...  &          ...  &    ...  &            ...  &    ...  &            $<$1.1  &      & 
 &...\\
 Cha I-8& OBR    	   &                  $<$47.7  &      & 361.0$\pm$       15.7 &  -14.3 &          $<$56.2  &      &101.0$\pm$        5.6 &   -4.6 & 49.6$\pm$        3.7 &   -3.6 & \ion{He}{I} & 3.0$\pm$0.3\\
 Cha I-9& --\,--\,R  &                  ...  &    ...  &            ...  &    ...  &          ...  &    ...  &            ...  &    ...  & 1.1$\pm$        0.2 &   -6.0 & 
 & ... \\
 Cha I-10& --\,--\,R  &                  ...  &    ...  &            ...  &    ...  &          ...  &    ...  &            ...  &    ...  &  1.7$\pm$        0.2 &   -3.8 & 
 & ... \\
 Cha I-11& --\,BR 	   &                  ...  &    ...  &             ...  &    ...  &          ...  &    ...  &  3.5$\pm$        0.4 &   -8.4 &  2.6$\pm$        0.5 &   -5.4 & 
 & 2.6$\pm$0.6\\
 Cha I-12& O--\,R 	   &        0.4$\pm$        0.1 &  -46.0 &   1.2$\pm$        0.1 &  -47.7 &1.2$\pm$        0.2 &  -10.9 &            ...  &    ...  &  2.6$\pm$        1.1 &   -2.6 & 
 & ...\\
 Cha I-13& OBR 	   &        7.6$\pm$        1.4 &   -1.0 & 557.0$\pm$        3.9 &  -67.2 &1.3$\pm$        3.2 &   -2.0 & 48.2$\pm$        3.3 &   -4.7 & 46.9$\pm$        4.7 &   -6.8 & \ion{He}{I}, [\ion{S}{II}] & 1.5$\pm$0.2\\
 Cha I-14& --\,--\,R  &                  ...  &    ...  &            ...  &    ...  &          ...  &    ...  &            ...  &    ...  &  3.8$\pm$        1.6 &   -2.7 & 
 & ...\\
 Cha I-15& --\,--\,R  &                  ...  &    ...  &            ...  &    ...  &          ...  &    ...  &            ...  &    ...  &            $<$0.57  &      & 
 & ...\\
 Cha I-16& OBR 	   &        2.3$\pm$        0.1 &   -6.3 &  41.3$\pm$        0.2 &  -89.8 &1.1$\pm$        0.1 &   -2.7 &  4.1$\pm$        0.3 &   -3.8 &  3.5$\pm$        0.4 &   -5.1 & \ion{He}{I}, [\ion{S}{II}] & 1.9$\pm$0.2\\
 Cha I-17& OBR 	   &        0.08$\pm$        0.02 &   -2.2 &   0.9$\pm$        0.1 &  -15.5 &          $<$0.27  &      &  1.2$\pm$        0.2 &   -1.6 &  3.6$\pm$        1.2 &   -3.5 & 
 & 0.9$\pm$0.3\\
 Cha I-18& --\,BR 	   &                  ...  &    ...  &             ...  &    ...  &          ...  &    ...  &  3.8$\pm$        0.3 &  -10.7 &  5.9$\pm$        1.0 &   -4.0 & \ion{He}{I}, \ion{Mg}{I} & 3.4$\pm$0.6\\
 Cha I-19& --\,BR 	   &                  ...  &    ...  &             ...  &    ...  &          ...  &    ...  &            $<$0.55  &      &  1.6$\pm$        0.4 &   -3.8 &  & $<$3.6\\
 Cha I-20& OBR 	   &                  $<$0.10  &      &   9.4$\pm$        0.1 & -107.2 &0.9$\pm$        0.1 &   -5.2 &  1.4$\pm$        0.2 &   -2.5 &  2.2$\pm$        0.4 &   -5.4 & 
 & 1.1$\pm$0.3\\
 Cha I-21& OBR 	   &        0.07$\pm$        0.02 &   -2.1 &   8.5$\pm$        0.1 & -181.5 &2.4$\pm$        0.1 &  -23.3 &  5.8$\pm$        0.1 &  -20.7 &  4.0$\pm$        0.2 &  -11.8 & \ion{He}{I}, \ion{Mg}{I} & 4.5$\pm$0.2\\
 Cha I-22& --\,BR 	   &                  ...  &    ...  &             ...  &    ...  &          ...  &    ...  & 50.7$\pm$       22.0 &   -8.5 & 37.0$\pm$        8.1 &   -3.4 & \ion{He}{I}, H$_2$ & 2.9$\pm$1.4\\
 Cha I-23& OBR 	   &        0.3$\pm$        0.1 &   -2.0 &  19.0$\pm$        0.1 &  -90.4 &0.5$\pm$        0.1 &   -1.7 &  1.1$\pm$        0.2 &   -1.1 &  4.1$\pm$        0.6 &   -4.9 & \ion{He}{I} & 0.6$\pm$0.1\\
 Cha I-24& --\,--\,R  &                  ...  &    ...  &            ...  &    ...  &          ...  &    ...  &            ...  &    ...  &  2.1$\pm$        0.5 &   -3.7 & 
 & ...\\
 Cha I-25& OBR 	   &                  $<$0.16  &      &   0.3$\pm$        0.1 &   -2.9 &          $<$0.37  &      &  0.9$\pm$        0.3 &   -1.4 &            $<$2.2  &      &  & $>$1.2\\
 Cha I-26& OBR 	   &                  $<$0.32  &      &   0.8$\pm$        0.1 &   -3.6 &          $<$0.61  &      &            $<$3.3  &      &  3.6$\pm$        1.6 &   -2.4 & \ion{He}{I}  & $<$2.5 \\
 Cha I-27& OBR 	   &        0.24$\pm$        0.07 &   -1.0 &  22.4$\pm$        0.2 &  -75.7 &2.2$\pm$        0.2 &   -6.2 &  2.2$\pm$        0.4 &   -1.8 &  4.1$\pm$        0.9 &   -4.6 & & 0.9$\pm$0.3\\
 Cha I-28& --\,--\,R  &                  ...  &    ...  &            ...  &    ...  &          ...  &    ...  &            ...  &    ...  &  0.5$\pm$        0.1 &   -3.8 & 
 & ...\\
 Cha I-29& OBR 	   &        3.9$\pm$        0.2 &   -5.6&  86.7$\pm$        0.9 &  -86.3 &          $<$1.2  &      & 13.2$\pm$        0.6 &   -6.1 &  6.7$\pm$        1.2 &   -5.6 & \ion{He}{I}  & 2.3$\pm$0.4\\
 Cha I-30& OBR 	   &                  $<$0.28 &      &  48.0$\pm$        0.2 &  -69.8 &3.9$\pm$        0.1 &   -9.5 & 19.5$\pm$        0.5 &  -14.1 &  7.1$\pm$        0.4&   -7.5 & \ion{He}{I} & 4.1$\pm$0.2\\
 
\hline
\end{tabular}
~\\
~\\
Notes.\\
$^{a}$: O: EFOSC2 grism\#16 optical spectrum; B: SofI blue grism spectrum; R: SofI red grism spectrum.\\
$^{b}$: \ion{He}{I} at 1.08\um, \ion{Mg}{I} at 1.5\um, \ion{Na}{I} at 2.2\um, [\ion{S}{II}] at 0.67\um, H$_2$ at 2.12\um, CO($\Delta\nu=2$) bandheads longward of 2.23\um.
\end{tiny}
\end{sidewaystable*}

\begin{sidewaystable*}
\begin{tiny}
\caption{Cha II sample: measured flux (not corrected for extinction) of the five emission lines used as accretion tracers (see caption of Table~\ref{tab:chai_lines}).}
\label{tab:chaii_lines}
\begin{tabular}{l|c|cc|cc|cc|cc|cc|c|c}
\hline
\hline
Id & Spectral & \multicolumn{2}{c|}{[\ion{O}{i}] 0.630$\mu$m}  & \multicolumn{2}{c|}{H$\alpha$}  & \multicolumn{2}{c|}{\ion{Ca}{ii} 0.854$\mu$m}  & \multicolumn{2}{c|}{Pa$\beta$}  & \multicolumn{2}{c|}{Br$\gamma$}  & other lines$^{b}$ & $R$\\ 
   &  segment$^{a}$ &(F $\pm$ $\Delta$ F)10$^{-14}$ & EW & (F $\pm$ $\Delta$ F)10$^{-14}$ & EW & (F $\pm$ $\Delta$ F)10$^{-14}$ & EW & (F $\pm$ $\Delta$ F)10$^{-14}$ & EW & (F $\pm$ $\Delta$ F)10$^{-14}$ & EW & &\\
   & &erg s$^{-1}$ cm$^{-2}$ & \AA & erg s$^{-1}$ cm$^{-2}$ & \AA & erg s$^{-1}$ cm$^{-2}$ & \AA & erg s$^{-1}$ cm$^{-2}$ & \AA & erg s$^{-1}$ cm$^{-2}$ & \AA & &\\
\hline

 Cha II-1& OBR 	   	   &        1.6$\pm$    0.1 &    -30.3  &  9.3 $\pm$   0.2 &    -135.8  & 16.9$\pm$        0.1 &    -53.9  & 44.3$\pm$        0.9 &    -16.9  & 150.0 $\pm$  5.8 &    -8.8  & \ion{He}{I}, H$_2$,CO, \ion{Na}{i}, \ion{Mg}{i}, [\ion{S}{ii}], [\ion{Fe}{ii}] & 1.9$\pm$0.1\\
 Cha II-2& OBR 	   	   &          $<$1.21          &      &   3.4$\pm$        0.2 &   -2.2 &           $<$3.6  &      &             $<$6.1  &      &  5.0$\pm$        1.0 &   -2.7 &  & $<$2.1\\
 Cha II-3& --\,--\,R   &                  ...  &    ...  &             ...  &   ...  &          ...  &    ...  &            ...  &    ...  &  1.4$\pm$        0.2 &   -5.9 & H$_2$
 & ...\\
 Cha II-4& --\,BR 	   &                 ...  &    ...  &            ...  &    ...  &          ...  &    ...  &  6.3$\pm$        0.3 &  -10.7 &  1.6$\pm$        0.3 &   -6.8 & \ion{Mg}{I} 
 & 5.8$\pm$1.0\\
 Cha II-5& OBR		   &                   $<$0.18  &      &   0.4$\pm$        0.1 &   -1.8 &          $<$0.50  &      &  0.9$\pm$        0.2 &   -1.1 &  2.6$\pm$        1. &   -3.7 & \ion{He}{I} & 0.7$\pm$0.3\\
 Cha II-6& OBR 	   	   &                   $<$0.58  &      &  49.5$\pm$        0.2 &  -66.0 &3.5$\pm$        0.5 &   -2.7 &  1.8$\pm$        0.3 &   -0.8 &  7.0$\pm$        1.1 &   -5.8 & 
 & 0.5$\pm$0.1\\
 Cha II-7& OBR 	       &                   $<$1.6  &      &  81.6$\pm$        0.4 &  -28.8 &3.2$\pm$        0.4 &   -1.6 &             $<$5.0  &      &  4.0$\pm$        0.7 &   -2.4 &  & $<$1.6\\
 Cha II-8& OBR 	   &                        $<$1.1  &      &  39.1$\pm$        0.2 &  -33.9 &           $<$1.2  &      &  3.0$\pm$        0.7 &   -2.5 &  1.8$\pm$        0.4 &   -3.4 & 
 & 1.8$\pm$0.6\\
 Cha II-9& OBR 	   &                        $<$1.4  &      &  12.9$\pm$        0.3 &   -2.5 &           $<$4.1  &      &  2.8$\pm$        0.5 &   -0.9 &  4.2$\pm$        1.3 &   -3.3 & 
 & 0.8$\pm$0.3\\
 Cha II-10& OBR 	   &                   $<$0.89  &      &   6.4$\pm$        0.5&   -3.1 &           $<$2.8  &      &             $<$4.0  &      &  3.9$\pm$        1.1 &   -3.5 &  & $<$1.6 \\
 Cha II-11& OBR 	   &        0.7$\pm$        0.1 &   -5.0 &  13.9$\pm$        0.2 &  -63.8 &1.3$\pm$        0.2 &   -3.4 &  3.8$\pm$        0.5 &   -5.9 &  1.8$\pm$        0.3 &   -3.9 & & 4.0$\pm$0.9\\
 Cha II-12& OBR 	   &        0.21$\pm$        0.07 &   -1.6 &   4.9$\pm$        0.1 &  -16.4 &           $<$0.92  &      &  1.9$\pm$        1.0 &   -2.4 &  0.8$\pm$        0.3 &   -2.9 & & 3.8$\pm$2.4\\
 Cha II-13& OBR 	   &        0.4$\pm$        0.1 &   -3.7 &   7.6$\pm$        0.2 &  -29.4 &           $<$2.0  &      &             $<$3.6  &      &  1.8$\pm$        0.4 &   -3.6 & \ion{He}{I} & $<$3.3\\
 Cha II-14& OBR 	   &                   $<$0.26  &      &  15.9$\pm$        0.1 &  -30.4 &2.8$\pm$        0.2 &   -4.1 &  2.5$\pm$        0.1 &   -1.7 &  2.7$\pm$        0.4 &   -2.1 & 
 & 1.8$\pm$0.3\\
 Cha II-15& OBR 	   &        1.3$\pm$        0.2 &   -1.5 & 127.0$\pm$        1.0 & -114.7 &5.1$\pm$        0.3 &   -3.7 & 31.0$\pm$        0.6 &   -8.1 & 13.8$\pm$        1.2 &   -5.0 & \ion{He}{I}  & 3.8$\pm$0.3\\
 Cha II-16& OBR 	   &                   $<$0.50  &      &   1.5$\pm$       -0.2 &   -1.1 &           $<$1.0  &      &  2.8$\pm$        0.4 &   -2.7 &  1.8$\pm$        0.4 &   -3.1 & 
 & 2.3$\pm$0.6\\
 Cha II-17& OBR 	   &        0.8$\pm$        0.1 &   -1.0 &  34.0$\pm$        0.2 &  -26.4 &1.8$\pm$       -0.3 &   -1.5 &             $<$7.2  &      &             $<$2.4  &      & 
 & ...\\
 
\hline
\end{tabular}
~\\
~\\
Notes.\\
$^{a}$: O: EFOSC2 grism\#16 optical spectrum; B: SofI Blue Grism spectrum; R: SofI Red Grism spectrum.\\
$^{b}$: \ion{He}{I} at 1.08\um, \ion{Mg}{I} at 1.5\um, \ion{Na}{I} at 2.2\um, [\ion{S}{II}] at 0.67\um, [\ion{Fe}{II}] at 1.25\um\, or 1.64\um, H$_2$ at 2.12\um, CO($\Delta\nu=2$) ro-vibrational bands longward of 2.23\um. 
\end{tiny}
\end{sidewaystable*}

\section{Determination of accretion properties}
\label{sec:accretion}

\subsection{Emission lines as accretion tracers}

\label{sec:tracers}
The wide spectral coverage of our data allows us to observe several emission lines both in the optical and 
near-IR that can be used to determine the accretion luminosity (\lacc) of the sources. 
The use of these lines as accretion tracers relies on the existence of empirical 
relationships connecting the line luminosity ($L_{line}$) to \lacc, which have been established by measuring 
the accretion luminosity through diagnostics that are independent of the emission lines, such as the UV/blue 
continuum excess \citep[e.g.][]{gullbring98,gullbring00,herczeg08}.
These relationships are typically written in the form
\begin{equation}
\mathrm{Log}\;L_{acc}/L_\odot = a \cdot \mathrm{Log}\;L_{line}/L_\odot + b~, 
\end{equation}
where $a$ and $b$ are parameters derived from fitting the observed luminosities
\citep[see e.g.][]{muzerolle98a,calvet00,calvet04,natta04,herczeg08,dahm08,fang09}. 
For our analysis we considered in particular five emission lines that are detected in many
objects of our sample: \oi\,$\lambda$6300, \caii\,$\lambda$8542, and the \hi\, transitions \ha, \pab, and \brg.
The line luminosity-\lacc\, relationships we employed are presented and briefly discussed
in the Appendix. We remark here that these relations are all substantially based on observations 
of young sources of the Taurus-Auriga complex.

The simultaneous detection of all or some of these five emission features allows us to 
derive accretion luminosities that are not biased by intrinsic line flux variations 
that may occur over long periods of time, as is the case when dealing with data 
acquired in different observing runs. 
In our survey the time span between observations of the same source with EFOSC2 and SofI has only been 24-48 hours, hence
our results are not biased by possible variations of the line flux on timescales longer than this interval.
Emission line variability on shorter timescales (hours) has been observed in several young objects: 
variations affect in particular the line profiles, but the
line flux changes at maximum by a factor of a few, which is much less than variations observed on longer timescales
\citep[e.g.][]{scholz05,stelzer07,mendigutia11}.

Apart from this limitation, the accretion luminosities derived from the five tracers should  
therefore be consistent with each other. This offers us the 
opportunity to compare the \lacc\, determinations and discuss the consistency and reliability 
of the different relationships employed.

In general, there will be a mix of mechanisms at work in different zones of the circumstellar region that contribute to the line formation,
which are more or less directly related to the ongoing accretion process. 
Indeed, there is evidence that some emission lines do not originate in accretion funnels (at least not only),
but rather in the disc, winds, and jets, i.e. they are indirect tracers of the accretion process.
For example, \oi\, is believed to form in winds/jets driven by accreting sources \citep[e.g.][]{cabrit90,hartigan95}. 
\caii\, emission has also been detected in jets \citep[e.g.][]{nisini05b,podio06}, indicating that part of the line might originate outside the magnetospheric
accretion region. Winds appear to strongly contribute also to \ha\, \citep[e.g.][]{reipurth96,calvet92b}.
Spatially extended emission suggesting a contribution from a wind is observed also for \hi\, infrared lines \citep[e.g.][]{beck10},
although in this case the emitting region is expected to be more compact with respect to the other tracers, because of the higher excitation conditions
of these lines, especially in the case of \brg.

Nevertheless, we point out that the line-\lacc\, relationships represent an \textit{empirical result}, 
so that they do not strictly depend on the actual physical mechanism regulating the accretion.
The comparison between the various \lacc\, 
determinations obtained in other samples than those over which the relations have 
been calibrated can therefore reveal particular trends and/or systematic biases
of these empirical formulas, thus indicating the tracers that are less affected by these problems.

\begin{figure}[!h]
\centering
\includegraphics[width=9.2cm]{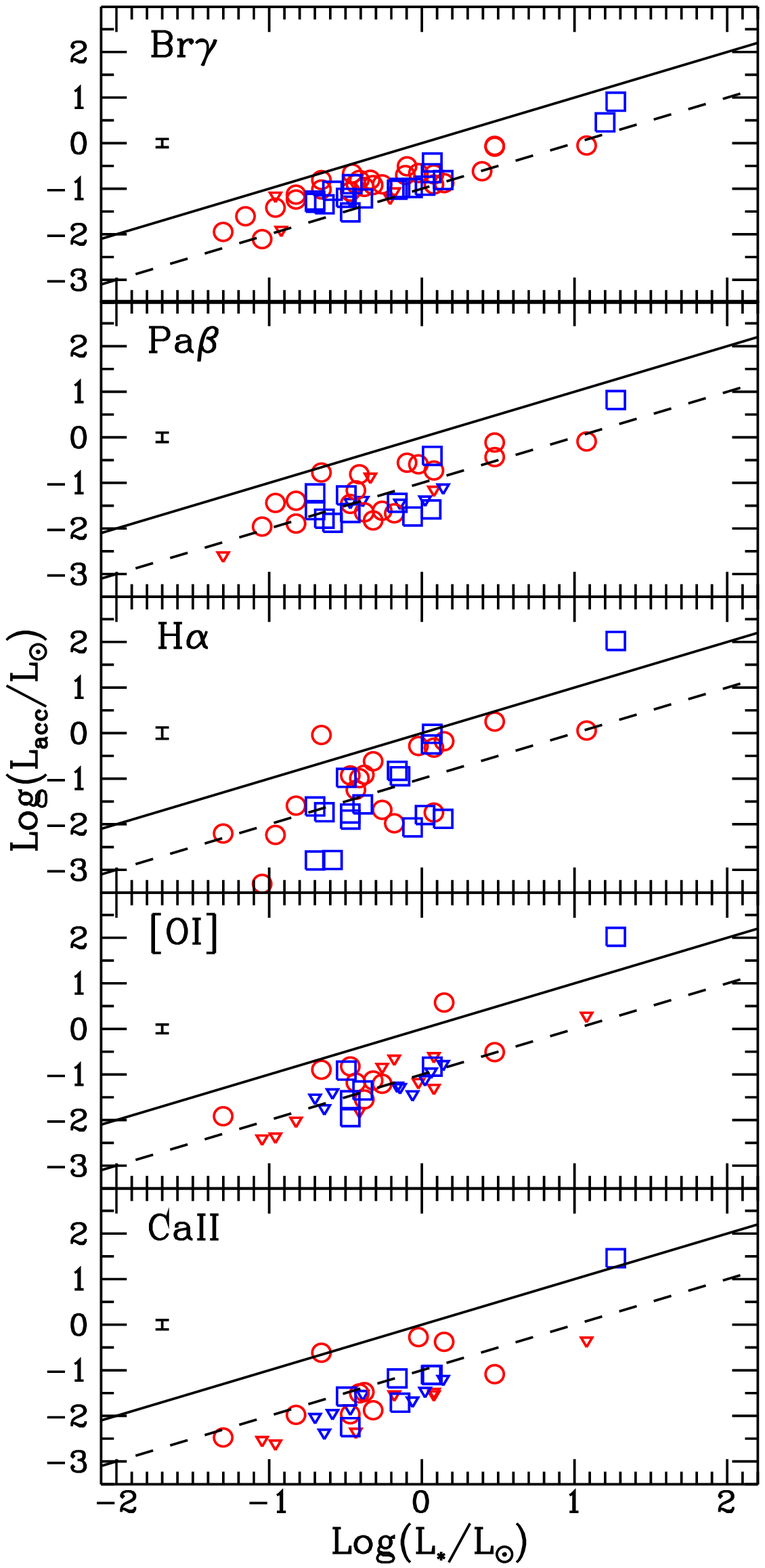}
\caption{\label{fig:laccs} \lacc\, values for the different accretion tracers considered 
(see Tables~\ref{tab:accretion1} and \ref{tab:accretion2}) plotted as a function of \lstar: from top 
to bottom Br$\gamma$, Pa$\beta$, H$\alpha$, \oi~$\lambda6300$, \caii~$\lambda8542$. Red circles and blue squares refer to detections of Cha I and Cha II objects, respectively.
Upper limits are indicated by downward triangles. The solid and dashed lines show the locus where \lacc=\lstar\, and \lacc=0.1\lstar, respectively.
The error bars reflect only the assumed uncertainty of 0.5 mag for the 2MASS $K$ band magnitude used as reference for the absolute flux calibration of the spectra 
(see Sect.~\ref{sec:observations}).
}
\end{figure}

\begin{sidewaystable*}
\begin{tiny}
\caption{Line luminosities of the five emission tracers considered and accretion luminosities derived using the relevant empirical relationships 
(see appendix) for Cha I objects. Upper limits are provided in case of non-detections. The shown mass accretion rates are computed from the the \brg\, 
accretion luminosity, using the available stellar parameters given in 
Table\ref{tab:chai_prop}.}
\label{tab:accretion1}
\begin{tabular}{l|ccccc|ccccc|c}
\hline
\hline
Id & $L_{\ion{[O]}{i}}$ & $L_{\ion{Ca}{ii}}$ & $L_{H\alpha}$ & $L_{Pa\beta}$ & $L_{Br\gamma}$ & \lacc$_{\ion{[O]}{i}}$ & \lacc$_{\ion{Ca}{ii}}$ & \lacc$_{H\alpha}$ & \lacc$_{Pa\beta}$ & \lacc$_{Br\gamma}$ & \macc$_{Br\gamma}$\\
   & \lsun       & \lsun         & \lsun       & \lsun        & \lsun   & \lsun       & \lsun & \lsun       & \lsun        & \lsun    &\msunyr     \\
\hline

Cha I-1    &  $<$5.39 $10^{-5}$  & $<$1.11 $10^{-4}$  &   8.49 $10^{-3}$  &   3.62 $10^{-4}$     &   1.04 $10^{-4}$     &  $<$5.04 $10^{-2}$  &  $<$2.94 $10^{-2}$  & 4.80 $10^{-1}$ &   1.87 $10^{-1}$      &  2.06 $10^{-1}$       &       1.96 $10^{-8}$  \\
Cha I-2    &     1.22 $10^{-5}$  &    1.33 $10^{-5}$  &   2.64 $10^{-4}$  &   $<$3.97 $10^{-6}$  &   4.10 $10^{-6}$     &  1.21 $10^{-2}$     &    3.36 $10^{-3}$   & 6.27 $10^{-3}$ &   $<$2.53 $10^{-3}$   &  1.13 $10^{-2}$       &       1.61 $10^{-9}$  \\
Cha I-3    &  $<$7.30 $10^{-5}$  &    1.91 $10^{-3}$  &   9.07 $10^{-3}$  &   4.05 $10^{-4}$     &   1.12 $10^{-4}$     &  $<$6.74 $10^{-2}$  &    5.34 $10^{-1}$   & 5.21 $10^{-1}$ &   2.56 $10^{-1}$      &  2.20 $10^{-1}$       &       1.42 $10^{-8}$  \\ 
Cha I-4    &  $<$3.76 $10^{-6}$  & $<$1.16 $10^{-5}$  &   3.45 $10^{-5}$  &   2.00 $10^{-5}$     &   2.74 $10^{-6}$     &  $<$3.91 $10^{-3}$  &  $<$2.93 $10^{-3}$  & 4.93 $10^{-4}$ &   1.10 $10^{-2}$      &  7.83 $10^{-3}$       &       1.17 $10^{-9}$  \\
Cha I-5    &  $<$4.17 $10^{-6}$  & $<$9.75 $10^{-6}$  &   2.50 $10^{-4}$  &   6.51 $10^{-5}$     &   1.60 $10^{-5}$     &  $<$4.32 $10^{-3}$  &  $<$2.45 $10^{-3}$  & 5.84 $10^{-3}$ &   3.63 $10^{-2}$      &  3.84 $10^{-2}$       &       7.55 $10^{-9}$  \\
Cha I-6    &          ...  &         ...  &        ...  &        ...     &   $<$2.76 $10^{-5}$  &  ...          &  ...          &      ... &        ...      &  $<$6.27 $10^{-2}$    &            ...  \\
Cha I-7    &          ...  &         ...  &        ...  &        ...     &   $<$3.11 $10^{-5}$  &  ...          &  ...          &      ... &        ...      &  $<$6.97 $10^{-2}$    &            ...  \\
Cha I-8    &  $<$2.40 $10^{-3}$  & $<$1.60 $10^{-3}$  &   1.69 $10^{-2}$  &   1.07 $10^{-3}$     &   5.25 $10^{-4}$     &  $<$1.93 $      $  &  $<$4.45 $10^{-1}$  & 1.14 $      $ &   8.11 $10^{-1}$      &  8.88 $10^{-1}$       &       5.98 $10^{-8}$  \\
Cha I-9    &          ...  &         ...  &        ...  &        ...     &   1.09 $10^{-4}$     &  ...          &  ...          &      ... &        ...      &  2.15 $10^{-1}$       &           ...  \\
Cha I-10   &          ...  &         ...  &        ...  &        ...     &   4.45 $10^{-5}$     &  ...          &  ...          &      ... &        ...      &  9.62 $10^{-2}$       &       7.72 $10^{-9}$  \\
Cha I-11   &          ...  &         ...  &        ...  &   4.51 $10^{-5}$     &   3.33 $10^{-5}$     &  ...          &  ...          &      ... &   4.06 $10^{-2}$      &  7.42 $10^{-2}$       &       7.17 $10^{-9}$  \\
Cha I-12   &     4.85 $10^{-3}$  &    1.52 $10^{-3}$  &   1.10 $10^{-2}$  &        ...     &   6.35 $10^{-5}$     &  3.79 $      $     &    4.21 $10^{-1}$   & 6.63 $10^{-1}$ &        ...      &  1.33 $10^{-1}$       &       1.41 $10^{-8}$  \\
Cha I-13   &     3.57 $10^{-4}$  &    3.05 $10^{-4}$  &   2.44 $10^{-2}$  &   5.05 $10^{-4}$     &   4.91 $10^{-4}$     &  3.09 $10^{-1}$     &    8.22 $10^{-2}$   & 1.79 $      $ &   3.69 $10^{-1}$      &  8.36 $10^{-1}$       &       1.67 $10^{-7}$  \\
Cha I-14   &          ...  &         ...  &        ...  &        ...     &   1.24 $10^{-4}$     &  ...          &  ...          &      ... &        ...      &  2.42 $10^{-1}$       &            ...  \\
Cha I-15   &          ...  &         ...  &        ...  &        ...     &   $<$4.56 $10^{-6}$  &  ...          &  ...          &      ... &        ...      &  $<$1.24 $10^{-2}$    &            ...  \\
Cha I-16   &     1.68 $10^{-4}$  &    4.18 $10^{-5}$  &   2.76 $10^{-3}$  &   4.63 $10^{-5}$     &   3.94 $10^{-5}$     &  1.50 $10^{-1}$     &   1.08 $10^{-2}$    & 1.18 $10^{-1}$ &   3.46 $10^{-2}$      &  8.62 $10^{-2}$       &       7.13 $10^{-9}$  \\
Cha I-17   &     6.75 $10^{-5}$  &   $<$5.66 $10^{-5}$  &   6.87 $10^{-4}$  &   1.98 $10^{-5}$     &   5.93 $10^{-5}$     &  6.26 $10^{-2}$     &  $<$1.47 $10^{-2}$     & 2.07 $10^{-2}$ &   2.45 $10^{-2}$      &  1.25 $10^{-1}$       &       1.58 $10^{-8}$  \\
Cha I-18   &          ...  &         ...  &        ...  &   1.06 $10^{-4}$     &   1.63 $10^{-4}$     &  ...          &  ...          &      ... &   2.77 $10^{-1}$      &  3.10 $10^{-1}$       &       4.74 $10^{-8}$  \\
Cha I-19   &          ...  &         ...  &        ...  &   $<$2.60 $10^{-5}$  &   7.67 $10^{-5}$     &  ...          &  ...          &      ... &   $<$1.35 $10^{-1}$   &  1.57 $10^{-1}$       &            ...  \\
Cha I-20   &  $<$9.56 $10^{-6}$  &    4.07 $10^{-5}$  &   8.14 $10^{-4}$  &   1.67 $10^{-5}$     &   2.53 $10^{-5}$     &  $<$9.58 $10^{-3}$  &  1.05 $10^{-2}$     & 2.56 $10^{-2}$ &   1.28 $10^{-2}$      &  5.78 $10^{-2}$       &       7.71 $10^{-9}$  \\
Cha I-21   &     1.42 $10^{-4}$  &    8.82 $10^{-4}$  &   1.42 $10^{-2}$  &   1.09 $10^{-4}$     &   7.54 $10^{-5}$     &  1.27 $10^{-1}$     &  2.42 $10^{-1}$     & 9.10 $10^{-1}$ &   1.69 $10^{-1}$      &  1.55 $10^{-1}$       &       2.36 $10^{-8}$  \\
Cha I-22   &          ...  &         ...  &        ...  &   7.06 $10^{-4}$     &   5.15 $10^{-4}$     &  ...          &  ...          &      ... &   7.69 $10^{-1}$      &  8.73 $10^{-1}$       &       1.42 $10^{-7}$  \\
Cha I-23   &     8.02 $10^{-5}$  &    5.12 $10^{-5}$  &   4.89 $10^{-3}$  &   1.55 $10^{-5}$     &   5.70 $10^{-5}$     &  7.38 $10^{-2}$     &  1.33 $10^{-2}$     & 2.41 $10^{-1}$ &   1.50 $10^{-2}$      &  1.20 $10^{-1}$       &       1.63 $10^{-8}$  \\
Cha I-24   &          ...  &         ...  &        ...  &        ...     &   9.98 $10^{-5}$     &  ...          &  ...          &      ... &        ...      &  1.99 $10^{-1}$       &            ...  \\
Cha I-25   &  $<$2.51 $10^{-4}$  & $<$1.11 $10^{-4}$  &   4.00 $10^{-4}$  &   1.56 $10^{-5}$     &   $<$3.96 $10^{-5}$  &  $<$2.21 $10^{-1}$  &  $<$2.92 $10^{-2}$  & 1.05 $10^{-2}$ &   2.15 $10^{-2}$      &  $<$8.67 $10^{-2}$    &    $<$7.91$10^{-9}$  \\
Cha I-26   &  $<$2.87 $10^{-4}$  & $<$1.26 $10^{-4}$  &   6.15 $10^{-4}$  &   $<$5.51 $10^{-5}$  &   5.87 $10^{-5}$     &  $<$2.51 $10^{-1}$  &  $<$3.33 $10^{-2}$  & 1.80 $10^{-2}$ &   $<$7.02 $10^{-2}$   &  1.23 $10^{-1}$       &       1.04 $10^{-8}$  \\
Cha I-27   &     2.96 $10^{-5}$  &    1.26 $10^{-4}$  &   2.84 $10^{-3}$  &   2.67 $10^{-5}$     &   5.13 $10^{-5}$     &  2.83 $10^{-2}$     &  3.33 $10^{-2}$     & 1.22 $10^{-1}$ &   2.26 $10^{-2}$      &  1.09 $10^{-1}$       &       1.41 $10^{-8}$  \\
Cha I-28   &          ...  &         ...  &        ...  &        ...     &   9.89 $10^{-6}$     &  ...          &  ...          &      ... &        ...      &  2.49 $10^{-2}$       &       4.09 $10^{-9}$  \\
Cha I-29   &     7.18 $10^{-5}$  & $<$1.76 $10^{-5}$  &   1.54 $10^{-3}$  &   1.20 $10^{-4}$     &   6.10 $10^{-5}$     &  6.64 $10^{-2}$     &  $<$4.46 $10^{-3}$  & 5.69 $10^{-2}$ &   6.87 $10^{-2}$      &  1.28 $10^{-1}$       &       1.55 $10^{-8}$  \\
Cha I-30   &  $<$1.58 $10^{-5}$  &    1.18 $10^{-4}$  &   2.48 $10^{-3}$  &   2.10 $10^{-4}$     &   7.61 $10^{-5}$     &  $<$1.55 $10^{-2}$  &  3.11 $10^{-2}$     & 1.03 $10^{-1}$ &   1.55 $10^{-1}$      &  1.56 $10^{-1}$       &       1.90 $10^{-8}$  \\

\hline
\end{tabular}
\end{tiny}
\end{sidewaystable*}

\begin{sidewaystable*}
\begin{tiny}
\caption{Line luminosities and derived accretion luminosities for Cha II objects (see caption of Table~\ref{tab:accretion1}. The mass accretion 
rates are computed from the the \brg\, accretion luminosity, using the available stellar parameters given in Table~\ref{tab:chaii_prop}.}
\label{tab:accretion2}
\begin{tabular}{l|ccccc|ccccc|c}
\hline
\hline
Id & $L_{\ion{[O]}{i}}$ & $L_{\ion{Ca}{ii}}$ & $L_{H\alpha}$ & $L_{Pa\beta}$ & $L_{Br\gamma}$ & \lacc$_{\ion{[O]}{i}}$ & \lacc$_{\ion{Ca}{ii}}$ & \lacc$_{H\alpha}$ & \lacc$_{Pa\beta}$ & \lacc$_{Br\gamma}$ & \macc$_{Br\gamma}$\\
   & \lsun       & \lsun         & \lsun       & \lsun        & \lsun   & \lsun       & \lsun & \lsun       & \lsun        & \lsun    &\msunyr     \\
\hline

Cha II-1   &     1.54 $10^{-1}$  &    9.51 $10^{-2}$  &   6.34 $10^{-1}$  &   1.81 $10^{-3}$     &   6.12 $10^{-3}$     & 1.05 $10^{2}$      &  2.87 $10^{1}$     & 1.05 $10^{2}$ &   6.58 $      $      &   8.10 $      $      &       4.49 $10^{-7}$  \\
Cha II-2   &  $<$1.89 $10^{-4}$  & $<$2.39 $10^{-4}$  &   4.77 $10^{-4}$  &   $<$9.19 $10^{-5}$  &   7.53 $10^{-5}$     & $<$1.68 $10^{-1}$   &  $<$6.39 $10^{-2}$  & 1.31 $10^{-2}$ &   $<$7.86 $10^{-2}$   &   1.55 $10^{-1}$      &       1.60 $10^{-8}$  \\
Cha II-3   &          ...  &         ...  &        ...  &        ...     &   1.92 $10^{-3}$     & ...           &  ...          &      ... &        ...      &   2.85 $      $      &       9.10 $10^{-7}$  \\
Cha II-4   &          ...  &         ...  &   7.87 $10^{-4}$  &   8.35 $10^{-5}$     &   2.11 $10^{-5}$     & ...           &  ...          & 2.46 $10^{-2}$ &   5.90 $10^{-2}$      &   4.91 $10^{-2}$      &       2.70 $10^{-9}$  \\
Cha II-5   &  $<$4.25 $10^{-5}$  & $<$4.41 $10^{-5}$  &   9.04 $10^{-5}$  &   1.50 $10^{-5}$     &   4.12 $10^{-5}$     & $<$4.01 $10^{-2}$   &  $<$1.14 $10^{-2}$  & 1.64 $10^{-3}$ &   1.33 $10^{-2}$      &   8.99 $10^{-2}$      &       6.42 $10^{-9}$  \\
Cha II-6   &  $<$1.27 $10^{-4}$  &    2.91 $10^{-4}$  &   9.65 $10^{-3}$  &   2.95 $10^{-5}$     &   1.11 $10^{-4}$     & $<$1.15 $10^{-1}$   &  7.83 $10^{-2}$     & 5.63 $10^{-1}$ &   2.62 $10^{-2}$      &   2.19 $10^{-1}$      &       5.42 $10^{-8}$  \\
Cha II-7   &  $<$5.60 $10^{-5}$  &    7.44 $10^{-5}$  &   2.67 $10^{-3}$  &   $<$5.94 $10^{-5}$  &   4.85 $10^{-5}$     & $<$5.23 $10^{-2}$   &  1.95 $10^{-2}$     & 1.13 $10^{-1}$ &   $<$3.64 $10^{-2}$   &   1.04 $10^{-1}$      &       6.45 $10^{-9}$  \\
Cha II-8   &  $<$1.84 $10^{-5}$  & $<$1.67 $10^{-5}$  &   6.24 $10^{-4}$  &   3.18 $10^{-5}$     &   1.93 $10^{-5}$     & $<$1.80 $10^{-2}$   &  $<$4.23 $10^{-3}$  & 1.84 $10^{-2}$ &   1.64 $10^{-2}$      &   4.53 $10^{-2}$      &       4.50 $10^{-9}$  \\
Cha II-9   &  $<$3.84 $10^{-5}$  & $<$8.19 $10^{-5}$  &   3.38 $10^{-4}$  &   3.20 $10^{-5}$     &   4.82 $10^{-5}$     & $<$3.64 $10^{-2}$   &  $<$2.14 $10^{-2}$  & 8.55 $10^{-3}$ &   1.83 $10^{-2}$      &   1.03 $10^{-1}$      &       4.73 $10^{-9}$  \\
Cha II-10  &  $<$8.49 $10^{-5}$  & $<$1.32 $10^{-4}$  &   5.58 $10^{-4}$  &   $<$5.63 $10^{-5}$  &   5.48 $10^{-5}$     & $<$7.80 $10^{-2}$   &  $<$3.49 $10^{-2}$  & 1.60 $10^{-2}$ &   $<$4.28 $10^{-2}$   &   1.16 $10^{-1}$      &       1.10 $10^{-8}$  \\
Cha II-11  &     1.35 $10^{-4}$  &    1.01 $10^{-4}$  &   2.50 $10^{-3}$  &   5.97 $10^{-5}$     &   2.78 $10^{-5}$     & 1.22 $10^{-1}$      &  2.65 $10^{-2}$     & 1.04 $10^{-1}$ &   5.33 $10^{-2}$      &   6.31 $10^{-2}$      &       4.68 $10^{-9}$  \\
Cha II-12  &     2.83 $10^{-5}$  & $<$5.54 $10^{-5}$  &   5.92 $10^{-4}$  &   2.74 $10^{-5}$     &   1.22 $10^{-5}$     & 2.72 $10^{-2}$      &  $<$1.44 $10^{-2}$  & 1.72 $10^{-2}$ &   2.19 $10^{-2}$      &   3.01 $10^{-2}$      &       7.16 $10^{-9}$  \\
Cha II-13  &     4.75 $10^{-5}$  & $<$1.14 $10^{-4}$  &   8.56 $10^{-4}$  &   $<$5.22 $10^{-5}$  &   2.68 $10^{-5}$     & 4.46 $10^{-2}$      &  $<$3.01 $10^{-2}$  & 2.73 $10^{-2}$ &   $<$4.19 $10^{-2}$   &   6.11 $10^{-2}$      &       3.60 $10^{-8}$  \\
Cha II-14  &  $<$6.05 $10^{-5}$  &    2.47 $10^{-4}$  &   3.33 $10^{-3}$  &   4.01 $10^{-5}$     &   4.38 $10^{-5}$     & $<$5.63 $10^{-2}$   &  6.63 $10^{-2}$     & 1.49 $10^{-1}$ &   3.65 $10^{-2}$      &   9.49 $10^{-2}$      &       4.53 $10^{-9}$  \\
Cha II-15  &     1.66 $10^{-4}$  &    2.96 $10^{-4}$  &   1.48 $10^{-2}$  &   4.55 $10^{-4}$     &   2.02 $10^{-4}$     & 1.48 $10^{-1}$      &  7.95 $10^{-2}$     & 9.63 $10^{-1}$ &   3.92 $10^{-1}$      &   3.76 $10^{-1}$      &       2.01 $10^{-8}$  \\
Cha II-16  &  $<$3.21 $10^{-5}$  & $<$3.63 $10^{-5}$  &   8.89 $10^{-5}$  &   3.66 $10^{-5}$     &   2.38 $10^{-5}$     & $<$3.06 $10^{-2}$   &  $<$9.36 $10^{-3}$  & 1.61 $10^{-3}$ &   2.52 $10^{-2}$      &   5.49 $10^{-2}$      &       3.02 $10^{-9}$  \\
Cha II-17  &     1.17 $10^{-5}$  &    2.21 $10^{-5}$  &   4.63 $10^{-4}$  &   $<$7.43 $10^{-5}$  &   $<$2.49 $10^{-5}$  & 1.16 $10^{-2}$      &  5.65 $10^{-3}$     & 1.27 $10^{-2}$ &   $<$3.78 $10^{-2}$   &   $<$5.71 $10^{-2}$   &    $<$5.49$10^{-9}$  \\

\hline
\end{tabular}
\\[0.8ex]
Notes.\\
$^{a}$: a more detailed analysis of ChaII-1 (DK Cha) is presented in \citet{garcia_lopez11}.\\ 
\end{tiny}
\end{sidewaystable*}

\subsection{Accretion luminosities from tracers: comparison and discussion}
\label{sec:lacc}

The line luminosities derived from the measured fluxes (correcting for extinction and distance) and the 
accretion luminosities obtained through the formulas given in the appendix 
are listed in Tables~\ref{tab:accretion1} and \ref{tab:accretion2}. Upper limits are provided for non-detections.

No flux correction was applied to take into account the stellar 
photospheric absorption for the lines considered,
because this contribution is usually negligible for late spectral types. 
The effect of this correction on \pab\, and \brg\, is examined 
in the case of Cha II-1 (spectral type F0) by \citet{garcia_lopez11}, who accordingly obtain
slightly higher accretion luminosities (9.5 and 8.1 \lsun, respectively).

First of all, we note that in many sources the tracers provide very different \lacc\, values. Indeed, we obtain 
accretion luminosities spanning up to two orders of magnitude in ChaI-12 and ChaII-1 for example. 
Apart from these considerable fluctuations regarding single sources, it is interesting 
to analyse the general differences of the results derived from the various tracers.
To this aim, we plotted in Fig.~\ref{fig:laccs} the \lacc\, obtained from the five lines as 
a function of \lstar. 

All tracers substantially indicate that \lacc\, increases with \lstar.
However, the plots clearly show that the accretion luminosities derived from the five relationships are 
actually characterised by very different scatters for given values of \lstar. 

In particular, the \brg\, line provides the least scattered \lacc\, values, which basically fall 
in the range 0.1\lstar$<$\lacc$<$\lstar\, for any \lstar, 
and accordingly result in a quite tight correlation between \lacc\, and \lstar.
\pab, \caii, and \oi\, accretion luminosities display a larger dispersion of about two orders 
of magnitude, which in the case of \ha\, goes up to more than three orders of magnitude.
Previous works using infrared \hi\, emission lines as tracers have not observed 
such a tight correlation between \lacc\, and \lstar\, as the one we get with \brg:
\citet{garcia_lopez06} have used \brg\, to derive \lacc\, in a sample of Herbig stars 
and have observed a dispersion of about two dex; \citet{natta06} found a scatter of about two orders of 
magnitude using \pab\, to compute \lacc\, in their low-mass Oph sample, which is similar to the \lacc(\pab)
dispersion we see in our data. 

Although in general we cannot expect young sources with similar \lstar\, to present comparable 
accretion luminosities, in this work we are analysing a homogeneous 
sample of objects of the same star-forming cloud that are probably in a similar evolutionary stage 
(almost all sources are Class IIs). Therefore, it seems totally 
reasonable to expect that targets with similar \lstar\, will likely display \lacc\, of the same order of magnitude.
In this sense, therefore, the \brg\, appears to be the most reliable tracer, because it provides the smallest 
scatter of \lacc\, values throughout the covered range of \lstar. 
In addition, the values of \lacc\, we derive from the \brg\, relationship, which are of the order 
of a fraction of the stellar luminosity ($\sim$0.1-1 \lstar), agree with what may be expected for typical Class II objects that are, by definition,
characterised by a bolometric luminosity dominated by the photospheric contribution.

We point out that the infrared tracers are also less affected by reddening, so any error on the extinction 
value should produce a smaller effect on the \lacc\, derived from \brg\, and \pab.
Indeed, the different observed scatters might be caused in principle (at least in part) by a wrong extinction estimate.
However, considering the fairly low values of $A_V$ characterising most of our sources 
(see Tables~\ref{tab:chai_prop} and \ref{tab:chaii_prop}),
even 100\% relative errors on this parameter are not sufficient to explain the very large scatters observed, 
especially for the optical lines.

As already mentioned, \brg\, is in general expected to be emitted in much more compact regions 
around the source with respect to the other tracers. 
Additionally, the fact that at least part of the line emission is indeed coming from infalling material 
(hence from a region very close to the source) has often been shown by the analysis of the \brg\, profiles \citep[e.g.][]{folha01}.
Thus, flux losses due to the signal actually falling within the slit are less important for 
\brg, even if part of the emission comes from a wind/jet. The smaller dispersion we observe 
is likely linked to this effect as well.

Based on the previous considerations, we therefore chose to adopt the accretion 
luminosities derived from \brg. 

\begin{figure*}[!t] 
\centering
\includegraphics[width=16cm]{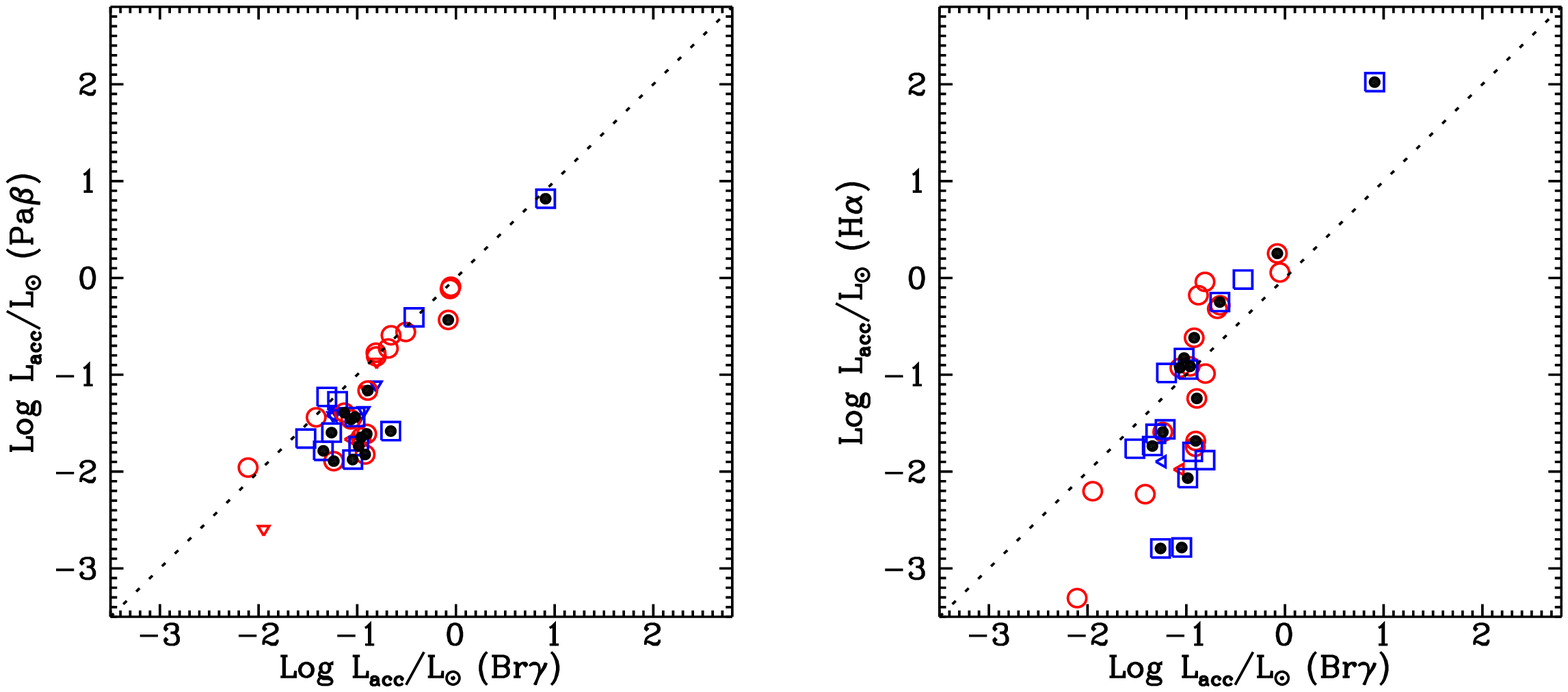}\\[-3ex]
\includegraphics[width=16cm]{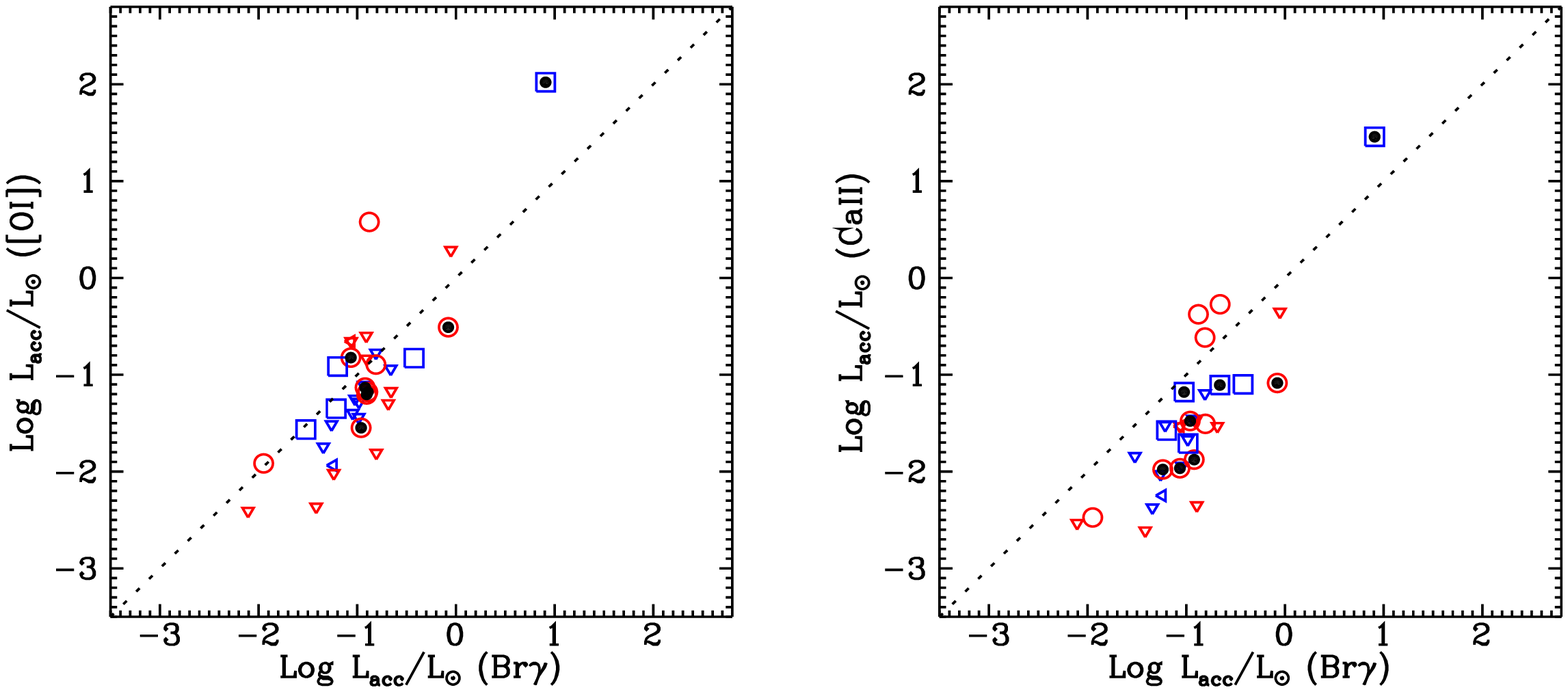}\\[-3ex]
\includegraphics[width=16cm]{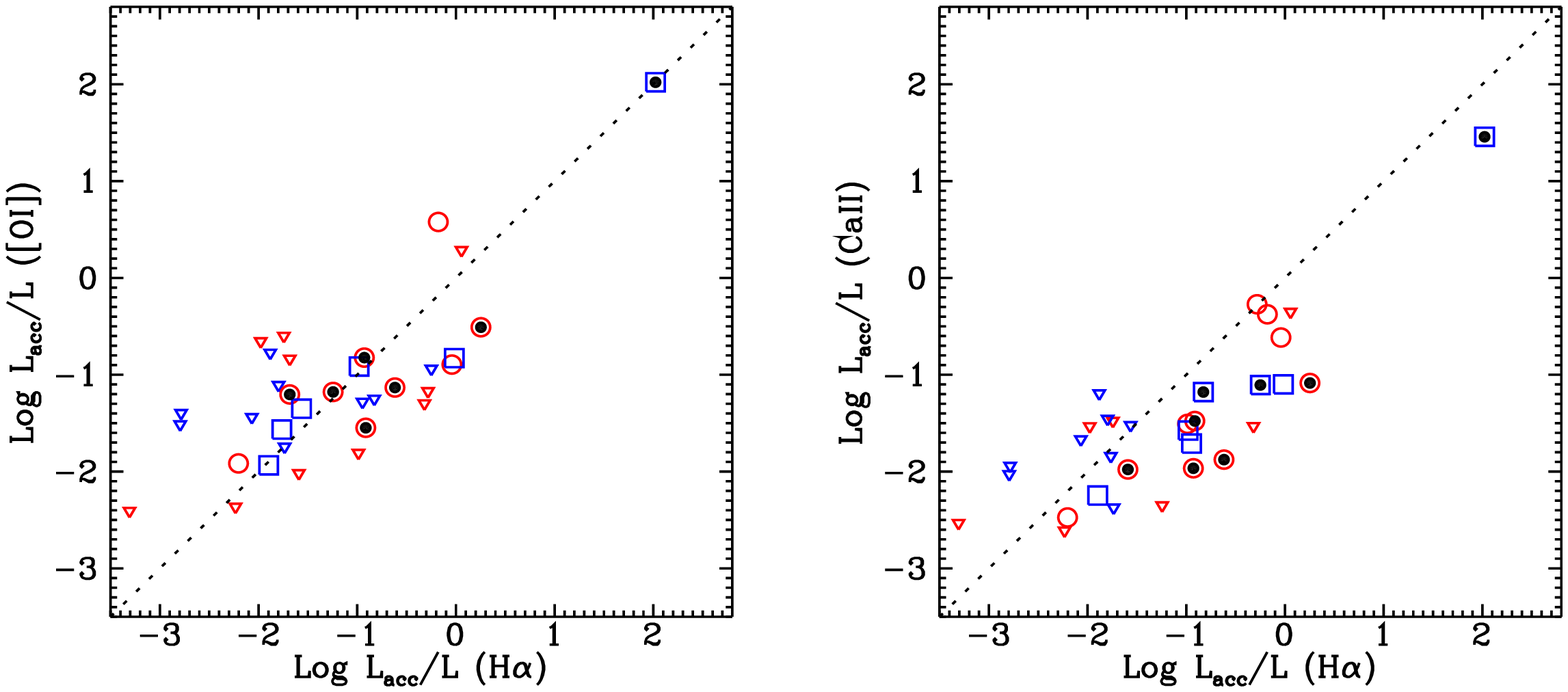}
\caption{\label{fig:comparisons} Comparison between \lacc(\brg) and \lacc\, determinations from the other four tracers 
and between \lacc(\ha) and \lacc\, from \oi\, and \caii.
The red circles and blue squares refer to Cha I and Cha II objects, respectively, whereas the triangles of the same colours indicate upper limits. 
The dashed lines mark the locus of equal accretion luminosity. 
The symbols filled with a black dot identify the sources displaying low values of the \pab/\brg\, ratio 
(see Sect.~\ref{sec:ratios} and Fig.~\ref{fig:ratios}).}
\end{figure*}

The larger scatters observed for the other tracers may probably be ascribed
to the different contributions to the emission lines, which may vary between our sources and those of the sample where the relationships 
were calibrated, for instance because of different line excitation mechanisms and/or different line formation
environments.
An important piece of evidence in favour of this scenario is that the other tracers 
provide mean accretion luminosities that are different from 
those obtained through \brg, in particular for some ranges of \lstar values.
To better analyse these discrepancies, we show in Fig.~\ref{fig:comparisons} direct 
comparisons between the accretion luminosities derived from \brg\, 
and those inferred from the other lines. 
We notice no particular difference between the distribution of Cha I and Cha II objects in these plots
and can consequently rule out effects related to specific properties of one of the two clouds.

There is evidence for \brg\, and \pab\, that in many cases \pab\, underestimates the 
accretion luminosity with respect to \brg\, in sources with $-2<$ Log \lacc(\pab) $<-1$, which 
corresponds to a range 0.1--1~\lsun\, for \lstar. 
In Sect.~\ref{sec:ratios} we show that these sources are those displaying unusually low values 
of the \pab/\brg\, ratio (see Fig.~\ref{fig:ratios}). In particular, these ratios are lower than 
those typically observed in the samples of T Tauri stars over which the \hi\, relationships were 
calibrated. 
Indeed, if we exclude these objects (indicated by filled symbols in Fig.~\ref{fig:comparisons}) 
the \brg\, and \pab\, accretion luminosity determinations appear to agree very well. 
Conversely, the low-\pab/\brg\, objects seem to be randomly distributed in the plots that show the comparison with the other tracers.

\lacc\, values from \brg\, and \ha\, are instead very different. Indeed, the \ha\, accretion luminosities seem to follow a 
different trend than \brg: they appear underestimated for \lacc(\brg)~$\lesssim$~0.1 \lsun\, and overestimated for \lacc(\brg)~$\gtrsim$~0.1 \lsun.
This effect is likely associated with the presence of several contributions to the line: 
for example enhanced chromospheric emission \citep[e.g.][]{hamann92} or emission/absorption from outflowing material,
the presence of which is often testified by P Cygni profiles \citep[e.g.][]{reipurth96,calvet92b}.
Also the contribution from direct photoionisation by stellar photons might become an important factor at higher luminosities. 
In all these scenarios, part of the \ha\, emission acts as an indirect tracer of the accretion process, and might be affected by strong
source-to-source variations.
For instance, if there is a significant contribution from a wind, then this emission is probably spatially extended
(much more than \brg), so that flux losses might become important, especially when using empirical relationships calibrated on the basis of 
spectroscopic observations carried out with different conditions and instrument configurations. 
This would explain in part the bigger scatter in \lacc\, values obtained from \ha.

We expect a similar situation for \oi\, as well, because forbidden oxygen emission is typically associated with outflowing winds/jets 
from protostars \citep[e.g.][]{ray07}. 
Indeed, if we take into account the upper limits on \lacc(\oi), we note that the shape of the correlation between \lacc\, values from \brg\, and \oi\,
looks similar to the one between \brg\, and \ha. 
A direct comparison between accretion luminosities derived from \ha\, and \oi\, shows that these tracers provide on average roughly consistent results, 
although the \lacc(\ha)-\lacc(\oi) correlation is
characterised by a large scatter around the locus of equal accretion luminosity (see Fig.~\ref{fig:comparisons}).
An evident correlation between \lacc(\ha) and \oi\, emission was also observed, for example, by \citet{fang09} in their sample of Orion YSOs.

Furthermore, the accretion luminosities derived from \caii\, seem to present the same trend as \oi\, and \ha\, when compared to \lacc(\brg), which might testify to 
a common origin of these lines from outflowing gas.
In particular, they appear to be underestimated with respect to \brg\, for \lacc~$\lesssim$~0.1 \lsun, which is similar to what is observed 
for \pab as well.
Interestingly, \lacc(\caii)\, appear to be correlated, but consistently (i.e. over the entire \lacc\, range) 
underestimated (about one dex) with respect to those from \ha.

In summary, our analysis shows that that the empirical line-\lacc\, relationships, calibrated on the basis of various surveys 
(but considering almost exclusively Taurus YSOs, see appendix), 
actually provide systematically different results when applied to our sample.
The \brg\, line gives the smallest dispersion of \lacc\, over the entire range of \lstar, whereas the other tracers, especially \ha, provide much more 
scattered \lacc\, results that are not expected for our homogeneous sample of targets.
In this sense, the \brg\, relationship is the most reliable, because it looks to be less subject to biases when applied to other YSO samples.

\section{Accretion properties of Cha I and Cha II objects}
\label{sec:properties}

\begin{figure}[!t] 
\centering
\includegraphics[width=8.5cm]{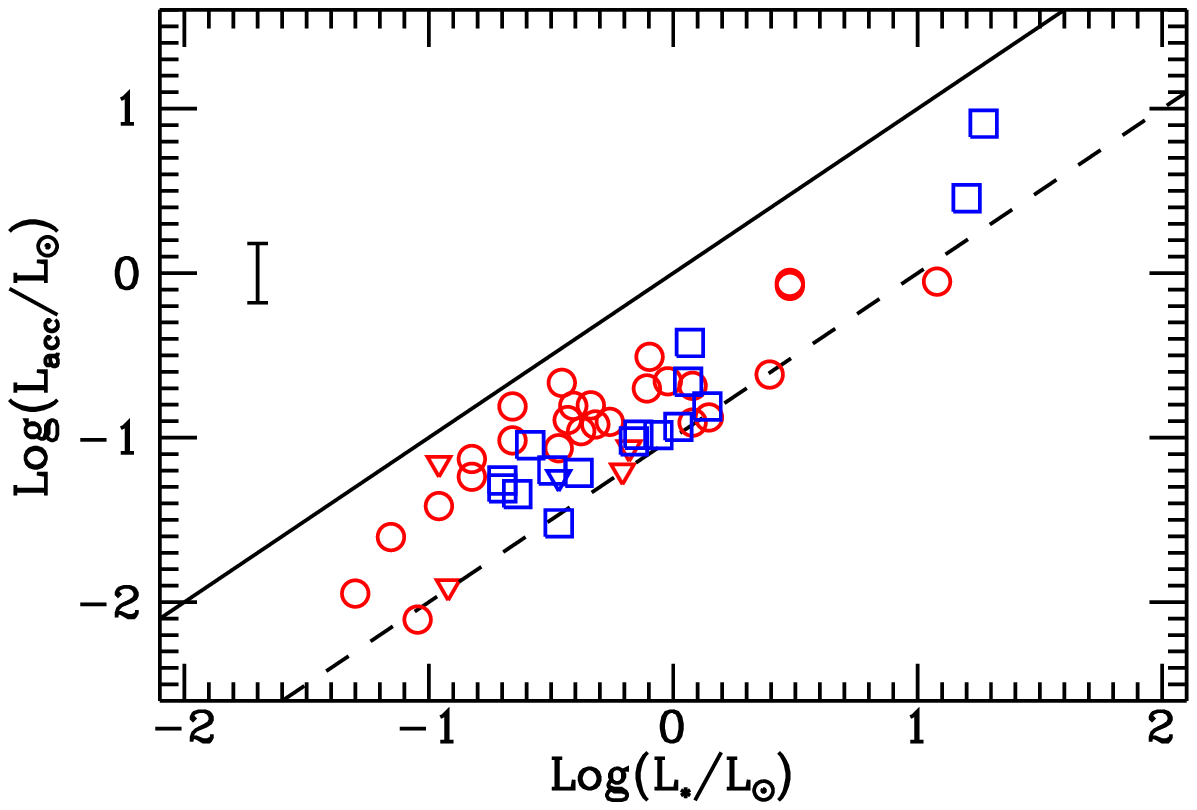}\\[-1ex]
\includegraphics[width=8.5cm]{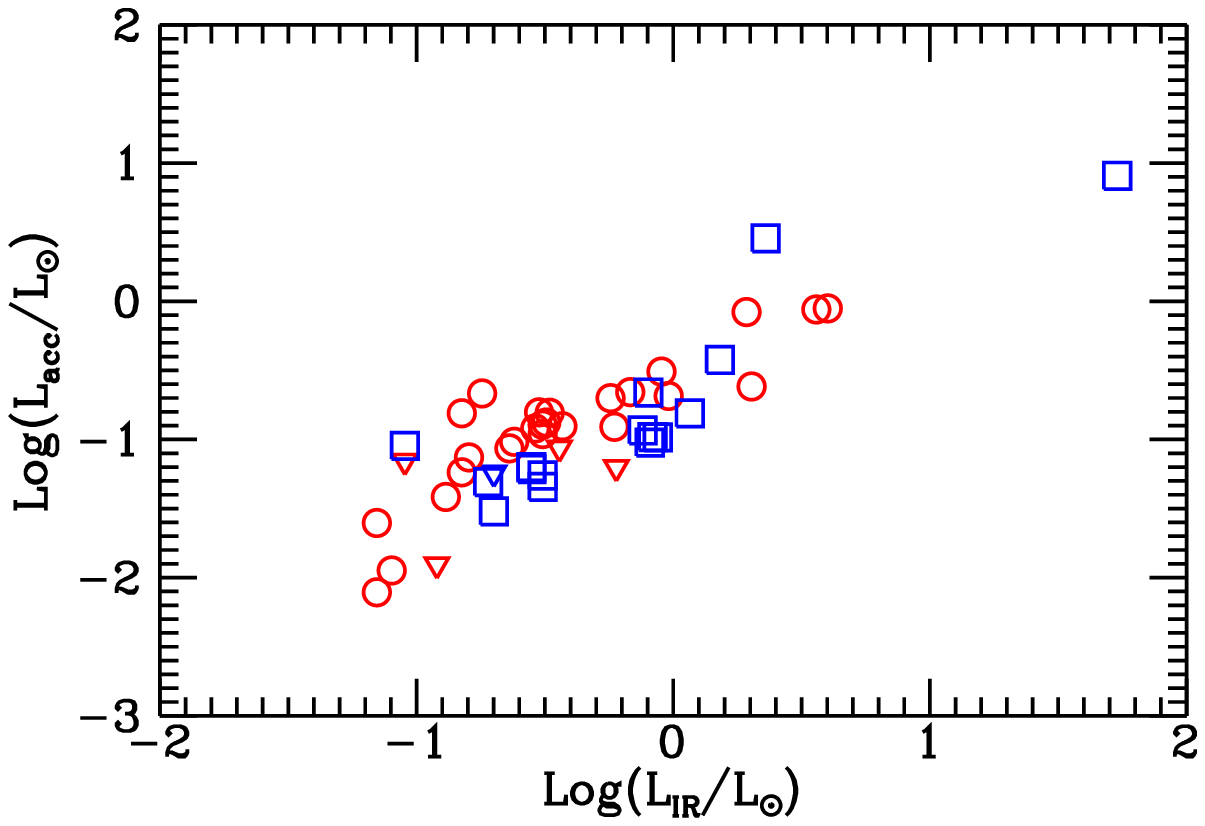}\\[-1ex]
\includegraphics[width=8.5cm]{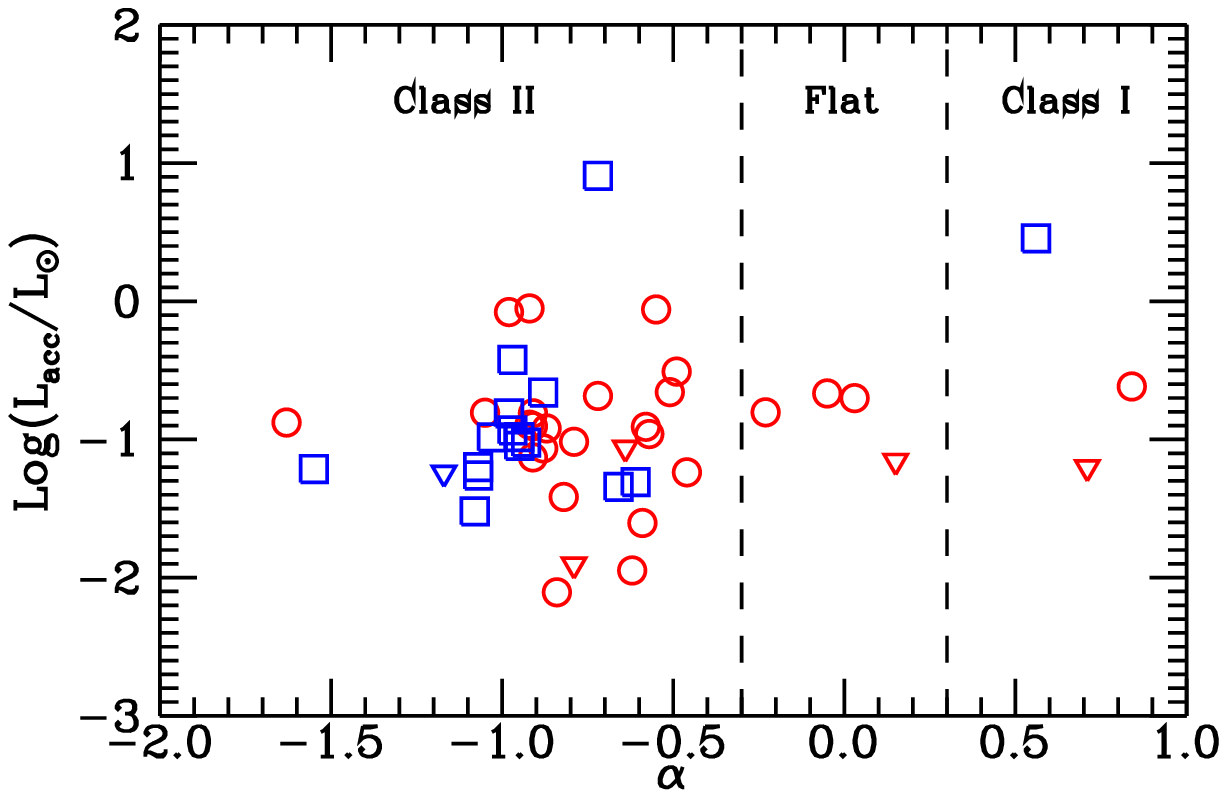}
\caption{\label{fig:extra_plots} Accretion luminosity (derived from \brg) as a function of \lstar (top), $L_{IR}$ (centre), 
and spectral index $\alpha_{2-24}$ (bottom).
Red circles and blue squares refer to Cha I and Cha II sources, respectively, while triangles of the same colours mark upper limits on \lacc.}
\end{figure}

In Fig.~\ref{fig:extra_plots} we analyse the connection between the accretion luminosity \lacc\, (derived from \brg) and other source parameters.
The top panel shows the tight correlation of \lacc\, and \lstar: all sources have accretion luminosities in the range 0.1~\lstar-1~\lstar, but
the two samples seem to display a different mean value of \lacc. We find indeed that the formal average accretion 
to stellar luminosity ratio for the Cha I and Cha II samples is \lacc/\lstar~=~0.29 and \lacc/\lstar~=~0.21, respectively.

Because the infrared luminosity $L_{IR}$ is very similar or comparable to \lstar\, for all sources (see Sect.~\ref{sec:parameters}), 
we find that \lacc(\brg) is also correlated to $L_{IR}$ (central panel). 
Conversely, we cannot derive any significant connection between \lacc\, and the YSO Class, also owing to the relative paucity of Flat and Class I sources. The bottom panel of Fig.~\ref{fig:extra_plots} shows that the Class II objects dominating the sample actually display
a very wide range of accretion luminosities.

\begin{figure}[!b] 
\centering
\includegraphics[width=8.5cm]{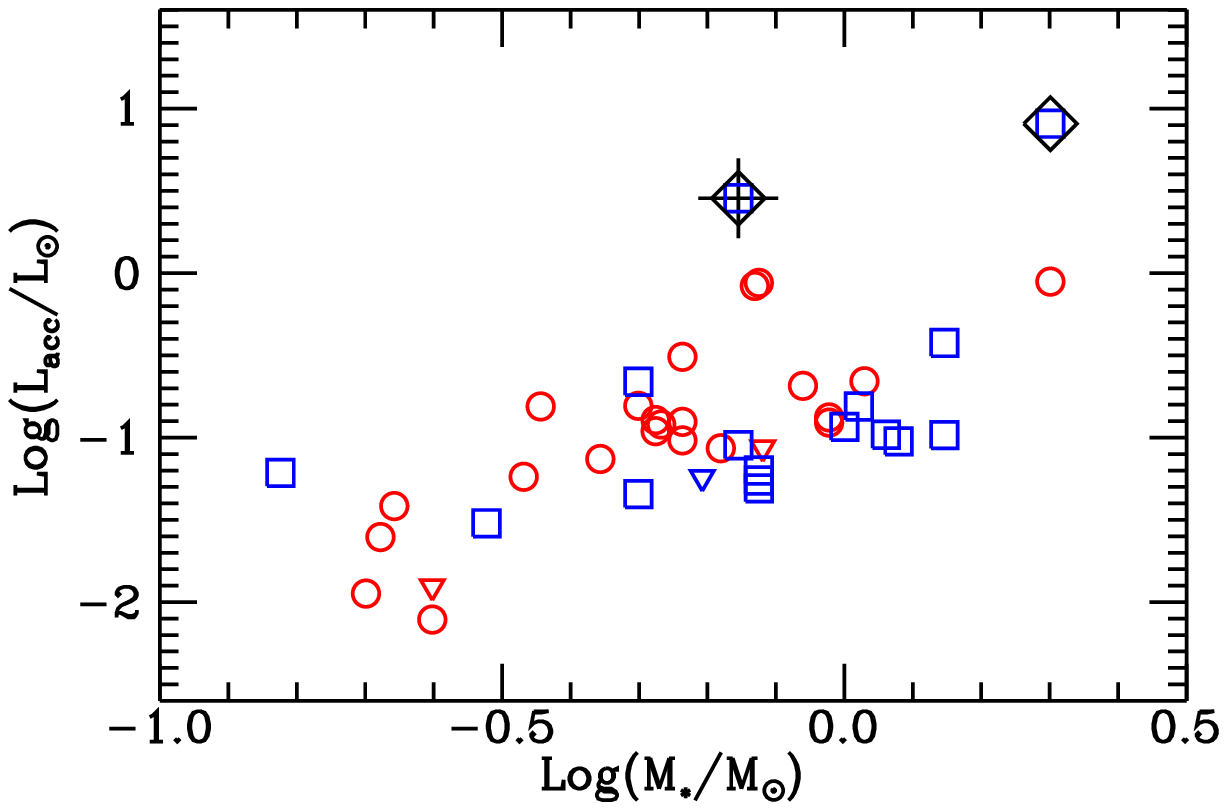}\\[-1ex]
\includegraphics[width=8.5cm]{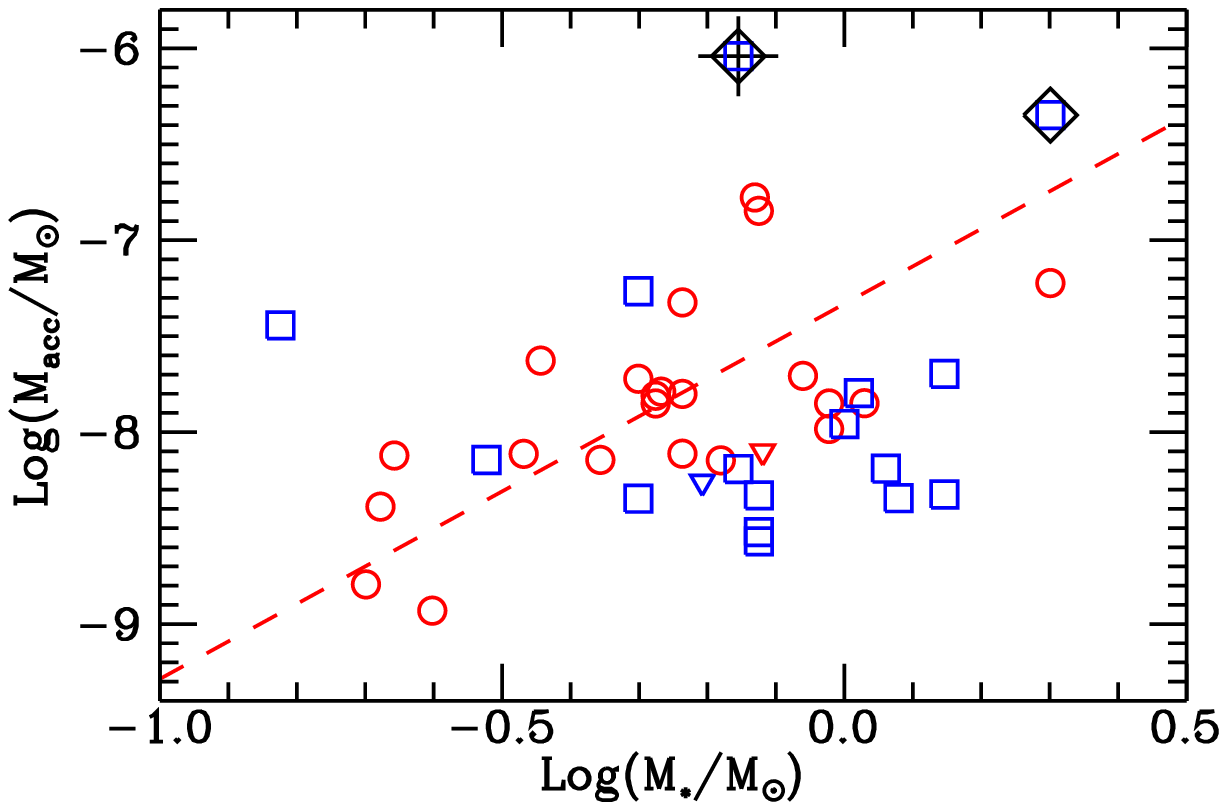}
\caption{\label{fig:macc_vs_mstar} \lacc\, (top panel) and \macc\, (bottom) as a function of the stellar mass $M_*$. Red circles 
and blue squares refer to Cha I and Cha II objects, respectively. Upper limits on \macc\, are marked by triangles of the same colours. 
The only Class I object (Cha II-3) is indicated by a cross. 
Diamonds identify the objects with a highly uncertain mass estimate (Cha II-1 and Cha II-3, see text for details). The dashed red line in the bottom panel shows the best-fit relation \macc $\varpropto M_{*}^{1.95\pm0.34}$ obtained for Cha I objects.}
\end{figure}

Since the \lstar of an object is determined by its mass and age, it is interesting to analyse how \lacc correlates with these quantities.
However, basically all sources of the sample have an estimated age in the range 1-10 Myr (except for three objects in Cha II with ages between 10 and 15 Myr), 
with a vast majority of objects being less than 5 Myr old \citep[see][]{luhman07,spezzi08}, i.e. substantially coeval. Consequently, the very narrow range of 
ages covered by the objects does not allow us to analyse the relation between accretion luminosity and age properly. 
The \lacc-\lstar\, correlation that we observe should therefore be ascribed mostly to a dependence of the accretion process on the stellar mass.
This \lacc-\mstar\, correlation is shown in the top panel of Fig.~\ref{fig:macc_vs_mstar}.

From \lacc(\brg) we computed the mass accretion rates \macc\, using \citep[e.g.][]{gullbring98}:
\begin{equation}
\dot{M}_{acc} = \frac{L_{acc}R_{*}}{GM_{*}} \left(1-\frac{R_{*}}{R_{in}}\right)^{-1} ~,
\end{equation}
where $M_{*}$ and $R_{*}$ are the known stellar parameters. We assume the factor $1-R_{*}/R_{in}$ equal to 0.8, which implies an inner disc radius $R_{in}= 5 R_{*}$.
The \macc\, obtained (see Tables \ref{tab:accretion1} and \ref{tab:accretion2}) are of the order of 10$^{-7}$-10$^{-9}$ \msunyr, in the range of values observed in 
many Class II objects \citep[see e.g.][]{gullbring98,natta06,white07}.

We have compared the derived mass accretion rates with those computed by \citet{spezzi08} for Cha II objects from the \ha\, 10\%-peak line width \citep{natta04}.
For almost all sources the \macc\, are lower (even two orders of magnitude in some cases) than the values we infer from \brg. 
Interestingly, in Fig.~\ref{fig:comparisons} (2nd panel) we find that the majority of Cha II sources are located in the region of the plot where \ha\, accretion
luminosities underestimate those from \brg. This finding might also suggest, albeit in a qualitative way, that the
\ha\, 10\% relation employed by the authors is affected by a trend similar to the one that was shown for the relationship between \ha\, flux and \lacc.

The computed mass accretion rates depend of course on the mass adopted for the sources. In general, the estimate of the mass will vary depending on the 
pre-main sequence evolutionary tracks considered. In our case, however, the masses adopted for Cha I and Cha II objects have been determined by \citet{luhman08a} and \citet{spezzi08}
using the same evolutionary models of \citet{baraffe98}, so that this should not introduce a systematic error in the \macc\, results of the two samples.
Spezzi et al. (2008) have actually investigated in their paper the effect of adopting different evolutionary models, providing also the mass estimates 
obtained from the models of \citet{dantona97} and \citet{palla99}. We checked how our results changed by using these different masses
for Cha II objects and found that although the mass values may differ even by a factor 2-3 in a few cases, there is no significant change in the general distribution 
of \macc.

The assumed mass estimate of Cha II-1 and Cha II-3 was obtained from the \citet{dantona97} evolutionary tracks, because \citet{baraffe98} models 
did not provide a clear result for these more luminous sources. Moreover, \citet{spezzi08} explicitly state that the stellar parameters provided 
for Cha II-3 (which is a Class I) must be taken with care.

In the lower panel of Fig.~\ref{fig:macc_vs_mstar} we show \macc\, as a function of the stellar mass. 
We remark that all the sources in the diagram are classified as Class IIs (i.e. they have a negative spectral index), except for Cha II-3. 
The plot shows a general visible trend of \macc\, increasing with the stellar mass, in agreement
with the expected correlation between accretion and \mstar. This is evident in particular for Cha I objects.
However, the observed spread of \macc\, for any value of \mstar, especially in the case Cha II sources, is larger than the corresponding 
dispersion in the \lacc-\lstar\, plot. 
We point out, however, that the sources displaying the highest mass accretion rates are Cha II-1 and Cha II-3, for which the result is likely biased because of the problems mentioned 
in deriving their stellar mass.
Indeed, we notice an evident increase of the spread passing from \lacc\ vs \lstar\ to \lacc\ vs \mstar\, and then to \macc\ vs \mstar\, (Figs.~\ref{fig:extra_plots} and \ref{fig:macc_vs_mstar}), 
i.e. introducing the stellar mass in the plotted quantities. Hence, this finding suggests that the larger dispersion observed in the \macc-\mstar\ plot is most probably 
the result of the (often large) uncertainties in the determination of the stellar masses.

Adopting a bisection linear regression on Cha I points, we derive a best-fit relationship \macc $\varpropto M_{*}^{1.95\pm0.34}$.
This power law index is close to the one found in other low-mass star-forming regions, e.g. 1.8 \citep[][in $\rho$~Oph]{natta04}, 1.87 \citep[][in Taurus]{herczeg08}, 
2.1 \citep[][using observations in various clouds]{muzerolle05}, 1.6 \citep[][in $\sigma$~Ori]{rigliaco11a}, whereas it significantly differs
from the 2.8-3.4 value measured for example in Orion by \citet{fang09}.
Because Cha II points are much more scattered in the plot, it is difficult to clearly identify a similar trend, so that we could not perform a reasonable linear fit.
Excluding Cha II-1 and Cha II-3, we see however that the majority of the Cha II objects provide accretion rates lower than in Cha I (for the same value of the stellar mass),
in agreement with the observed difference of the average \lacc/\lstar values of the two clouds.
This small difference in the \macc\, value between Cha I and Cha II, might actually reflect the difference between the mean age of the two associations (i.e. $\sim$1-2 Myr, 
see Sect. \ref{sec:parameters}), suggesting a decrease of the mean accretion rate with time.


\section{Pa$\beta$/Br$\gamma$ ratio}
\label{sec:ratios}

\begin{figure*}[]  
\centering
\includegraphics[width=15cm]{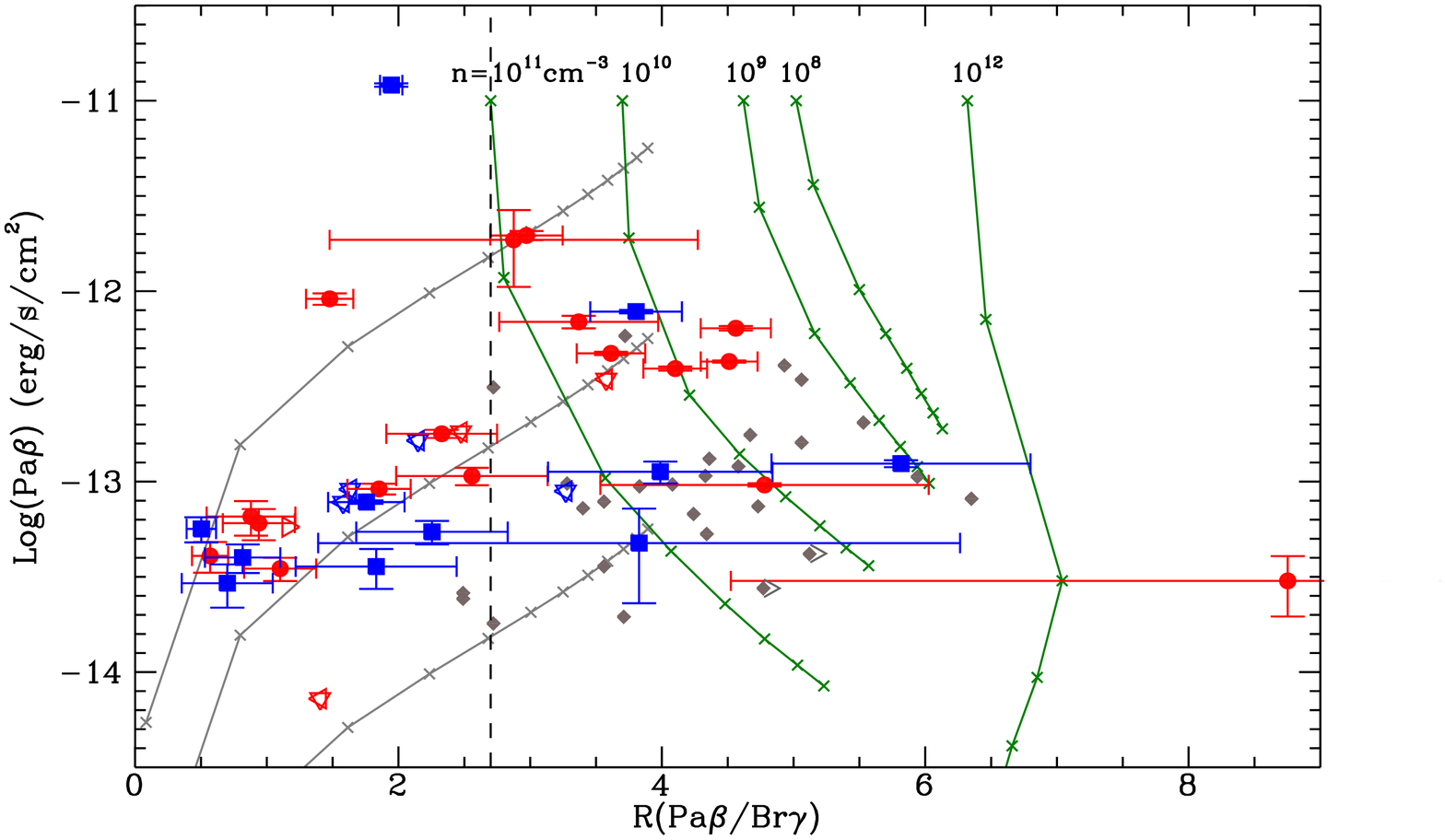}\\[-4ex]
\includegraphics[width=15cm]{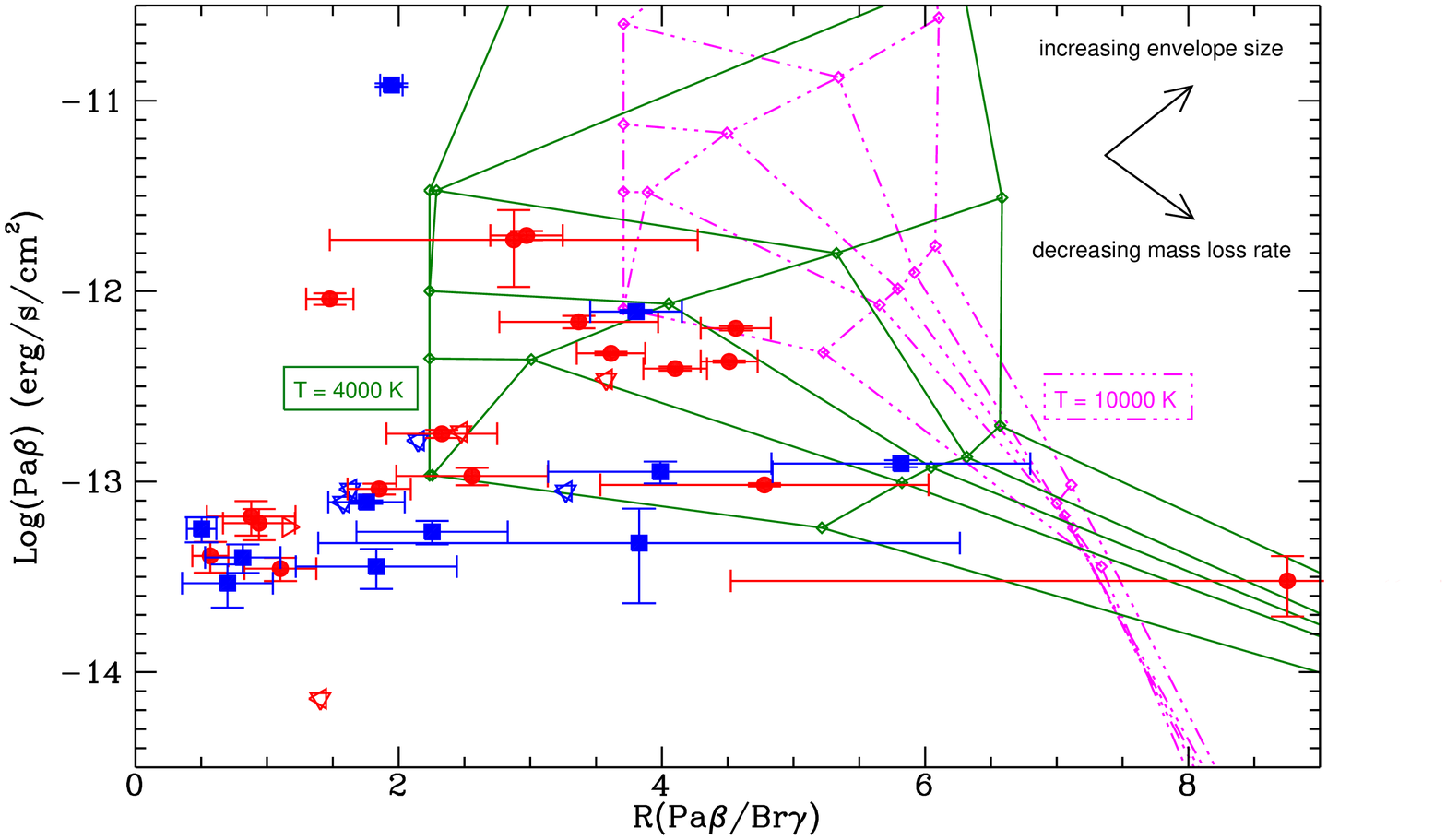}
\caption{\label{fig:ratios} \pab\, flux as a function of the Pa$\beta$/Br$\gamma$ ratio for ChaI (red points) and ChaII (blue points) sources. The open triangles 
indicate upper/lower limits, depending on their orientation. 
All quantities are extinction-corrected. A comparison with various emission scenarios is presented (see text for details).  
(\textit{Upper panel}) The grey solid lines show the expected emission from gas in an optically thick region in LTE and refer to three different emitting areas 
(0.2, 2, 20 $\times10^{22}$ cm$^2$) and temperatures (crosses) from 1000 to 12000 K (from left to right in steps of 1000 K). 
The green curves on the right side of the plot show the expected emission in case-B conditions from a gas with density $n$, indicated at the top of each curve, 
and temperatures (crosses) from 500 to 15000 K (downwards, steps corresponding to 500, 1000, 3000, 5000, 7500, 10000, 12500, and 15000 K).
The case B curves have been normalised to obtain the same \pab\, emission at $T=500$ K. The vertical dashed line at $R\sim2.7$ marks the lower limit 
of the region where ratios are still compatible with optically thin emission in case B conditions. For comparison's sake the grey diamonds show the ratios 
measured in Taurus CTTSs by \citet{muzerolle98b}.
(\textit{Lower panel}) Results obtained from a model of a spherically symmetric expanding wind in LTE conditions 
\citep[see][]{nisini95} for gas temperatures of T=4000 K (green) and T=10000 K (magenta) are overplotted. 
For each temperature the open diamonds show the expected emission for four different values of the mass loss rate (\mloss\, from $10^{-6}$ to $10^{-9}$ \msunyr) and envelope 
thickness (1,2,5,10, and 100 \rstar), these parameters varying in the direction indicated by the arrows. The lines (solid green for T=4000 K and 
dotted-dashed magenta for T=10000 K) connect the results obtained with the same thickness or \mloss.}
\end{figure*}

The analysis of the (extinction-corrected) Pa$\beta$/Br$\gamma$ ratio (hereafter $R$) provides important information on the regions and mechanisms 
responsible for the \hi\, emission, because this ratio is sensitive to the temperature and density of the emitting gas.
We find that the discrepancies between \lacc(\brg) and \lacc(\pab) observed in many objects (see first panel of Fig.~\ref{fig:comparisons}) can be related to 
a different intrinsic value of the Pa$\beta$/Br$\gamma$ ratio in these sources with respect to the ratio measured in the sample where the \hi\, 
relationships were calibrated.

In Fig.~\ref{fig:ratios} we show $R$ as a function of the de-reddened \pab\, flux. 
We notice that the sources of our samples display quite a wide range of $R$ values (from $\sim$0.5 up to $\sim$9), with no significant distribution 
difference between Cha I and Cha II objects.  
Remarkably, the figure also shows that there is a large group of sources (15 out of 28) presenting low $(\lesssim~2.5)$ $R$ values and \pab\, intrinsic fluxes of the order of
$10^{-14}$-$10^{-13}$ erg~s$^{-1}$~cm$^{-2}$, which appears to be somewhat separated from the rest of the objects that display intermediate-high ratios.

The position of the sources on the diagram depends of course on the extinction value, so we have analysed the effects of possible errors on the adopted $A_\mathrm{V}$, 
especially in regard to the group of objects showing low ratios. 
For instance, considering an $A_\mathrm{V}$ variation of $+4$ mag (i.e. adopted extinction is underestimated), we obtain an increase of both the \pab\, intrinsic flux 
(by a factor $\sim$3.4) and $R$ (factor $\sim$2.1).
Because most of the sources displaying low ratios have extinction estimates lower than 4 mag, this variation would represent a relative error greater than 100\% on the 
extinction. In any case, even assuming such a scenario, the sources now displaying $R\lesssim1$ would remain in the low ratio region of the diagram. 
Conversely, if the adopted extinction were overestimated, the derived $R$ value would be even lower.
At any rate, the quite uniform distribution of both Cha I and Cha II objects in the different regions of the diagram suggests that there is no 
relevant systematic error between the extinction estimates of the two samples. 
Additionally, we point out that because $R$ depends only on the goodness of the inter-calibration of the BG and RG segments, it does not suffer from the uncertainty 
on the absolute flux scale  calibration of the spectra, whereas this affects the intrinsic \pab\, flux. 

Now, we can show that different values of $R$ can be actually associated with different regimes of emission. This can be easily seen in the upper 
panel of Fig.~\ref{fig:ratios}, where we display the 
expected loci for two standard emission scenarios: optically thick lines from a gas in LTE conditions, i.e. blackbody-like emission, and case-B emission \citep{hummer87},
i.e. optically thin lines when level populations are mostly determined by radiative cascade from the continuum. 

The blackbody curves on the left side of the plot (solid grey lines) were computed assuming a FWHM of 100 km s$^{-1}$ (lines are not resolved in our spectra) 
and considering different projected areas for the emission region: (bottom to top) $2\times 10^{21}$, $2\times 10^{22}$, and $2\times 10^{23}$ cm$^{2}$.
To simplify the diagram we have traced these curves assuming a single average value for the distance to the Cha clouds: because the distances are very similar, 
this does not significantly affect their location on the plot.
The crosses along each line refer to different values of the gas temperature increasing from 1000 K up to 12000 (left to right), in steps of 1000 K. 
We note that the curves cover a range of $R$ spanning from very low values up to about 4.

The case-B curves on the right side of the plot (solid green lines) show the expected emission from a gas with density $n$ (from 10$^8$ to 10$^{12}$ cm$^{-3}$) 
within a given emitting volume $V$. The crosses
along the curves indicate different values of the gas temperature from 500 K up to 15000 (top to bottom). 
The various lines have been normalised so as to obtain the 
same \pab\, emission (which is proportional to the $n^{2}V$ product) at $T=500$K. We note that for case-B conditions the lowest values of $R$ (around 2.7) 
are obtained for gas densities of the order of 10$^{11}$ cm$^{-3}$ (and low temperatures), after which $R$ starts to increase with increasing density
(see the location of the 10$^{12}$ cm$^{-3}$ density curve). 

Therefore, although the expected \pab\, flux may vary in the two cases depending on the assumed emission area or $n^{2}V$ product, evidently 
low values of $R$ ($\lesssim 2.5$) are obtainable only for optically thick and 
thermalised lines
for the reasonable range of temperatures explored, whereas high ratios ($\gtrsim 4$) indicate optically thin lines.

Hence, the low $R$ values observed in many objects provide evidence that the \pab\, and \brg\, lines originate in an LTE gas at temperatures $1000 K < T \lesssim 4000-5000 K$. 
In this scenario, the size of the emitting (projected) area can be estimated directly 
from the position of the points with respect to the blackbody curves. The majority of the objects is clearly located in the region between the $2\times 10^{22}$ cm$^2$ 
and $2\times 10^{23}$ cm$^2$ curves or around them. This indicates an emitting region that is about 0.6-6 times the projected area of a young star with a typical 
radius of 2\,R$_\odot$.

In addition to the standard emission scenarios considered above, we have also taken into account the case of emission 
from a spherically symmetric expanding wind in LTE conditions. 
In the simple model we employed \citep[see][for a comprehensive description]{nisini95}, we assume a fixed gas temperature, an envelope inner radius
$r_{i}$~=1~\rstar, and a radial velocity law ($v \varpropto (1-r_i/r)^{0.2}$). The computation was then performed using Sobolev's large velocity gradient approximation \citep{sobolev60}. 
The lower panel of Fig.~\ref{fig:ratios} shows the 
results obtained from the model for two different gas temperatures (4000 K and 10000 K) and for a range of mass loss rates \mloss\, ($10^{-6}$--$10^{-9}$ 
\msunyr) and envelope thickness (1,2,5,10, and 100 \rstar).

In these models the line emission will be optically thick or thin depending on the assumed parameters. In any case, the locus 
of the results stretches from the central-upper part of the plot (intermediate values of $R$) towards the right-lower end of it (i.e. high $R$ values 
and decreasing \pab\, flux). 
Albeit in a qualitative way, we can see that the sources displaying intermediate-high Pa$\beta$/Br$\gamma$ ratios 
are located in the region compatible with the results of the wind model with a temperature of 4000 K, in positions corresponding to mass loss 
rates of the order of $10^{-8}$--$10^{-9}$ \msunyr.
Actually, the same region of the $R$-Log(\pab) plot can be covered also by the predictions of magnetospheric accretion models \citep[see][]{muzerolle01}, although
the expected gas temperatures are higher in this case ($T\sim6000-12000K$).

In summary, our analysis of the Pa$\beta$/Br$\gamma$ ratio clearly shows the presence of at least two different \hi\, emission modalities, 
one present in the objects showing low $R$ values and indicating
optically thick lines, the other related to intermediate/high $R$ ratios compatible with optically thin emission. 
Any calibration of the empirical line-\lacc\, relationships of \brg\, and \pab\, is clearly affected by the existence of these different emission modalities 
as shown by the \brg-\pab\, comparison in Fig.~\ref{fig:comparisons}, consequently it is important to understand their origin.

Observations of low $R$ values compatible with thermalised lines from small regions with gas at quite low temperatures have been reported also by \citet{gatti06} 
for many objects of their Ophiuchus sample, which is composed of low mass stars and brown dwarfs (BDs).

In this respect, it is interesting to compare our Fig.~\ref{fig:ratios} with the analogous Fig.~7 of \citet{gatti06}. 
We note that for the Cha sources the intrinsic \pab\, flux of low-ratio objects appears to be higher ($10^{-14}$-$10^{-13}$ erg s$^{-1}$ cm$^{-2}$) than 
in Ophiuchus BDs ($10^{-15}$-$10^{-14}$ erg s$^{-1}$ cm$^{-2}$).
Fig.~7 of \citet{gatti06} also includes the results found by \citet{muzerolle98a} for the T Tauri stars in Taurus-Auriga, which we report in Fig.\ref{fig:ratios} as well.
Remarkably, all sources of the Taurus-Auriga sample display high Pa$\beta$/Br$\gamma$ ratios ($R \gtrsim 3$). 
Their position on the diagram is similar to the one of the Cha sources displaying intermediate-high values of $R$, i.e. these objects seem to present the same modality 
for the \hi\, emission. Consistently with Fig.~\ref{fig:ratios}, \citet{bary08}, who have extensively analysed the Brackett 
and Paschen decrements measured in the Taurus-Auriga CTTSs, conclude for example that \hi\, 
emission is compatible with optically thin case-B recombination in a gas at very low temperature ($T\sim2000$K) and with electron densities of the 
order of $10^{10}$~cm$^{-3}$.

\begin{figure}
\centering
\includegraphics[width=9.0cm]{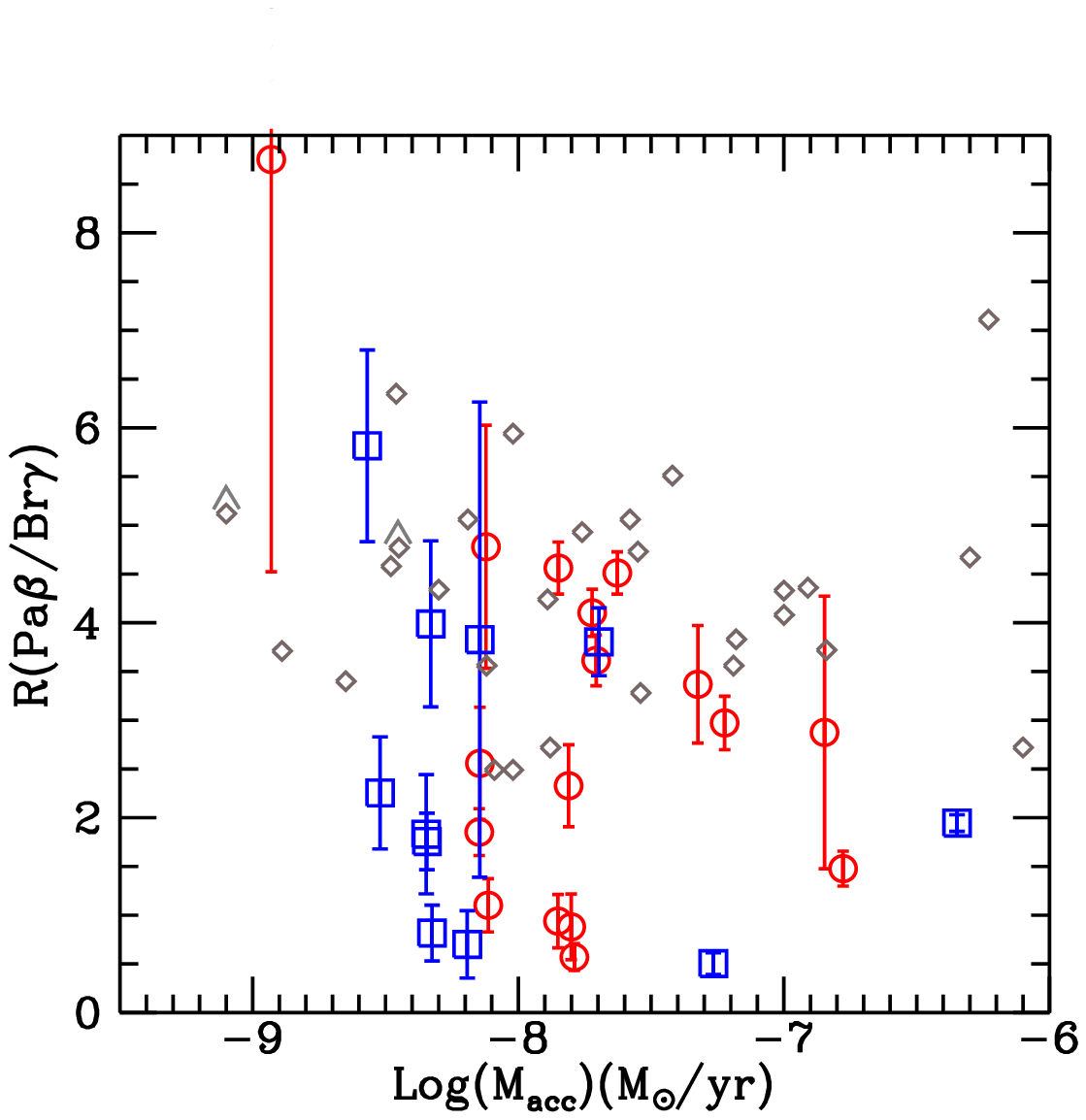}
\caption{\label{fig:ratios2} \macc\, as a function of the Pa$\beta$/Br$\gamma$ ratio. Red circles and blue squares refer to Cha I and Cha II
 sources, respectively. Measurements relative to Taurus CTTSs by \citet{muzerolle98a} are reported for comparison (grey diamonds). 
 }
\end{figure}

We find that the majority of Taurus objects display fluxes of the order of $10^{-14}$-$10^{-13}$~erg s$^{-1}$ cm$^{-2}$, like most of our Cha sources.
We also observe a few Cha objects that display \pab\, fluxes on average higher ($\sim10^{-13}$-$10^{-12}$ erg s$^{-1}$ cm$^{-2}$, ten objects
in the whole sample).
However, there is no clear separation in terms of mass accretion rates between these objects and the others (see for example the comparison with 
Taurus sources in Fig.~\ref{fig:ratios2}, which show \macc\, values that uniformly span three orders of magnitude).
Because we expect that the accretion luminosity is proportional to \pab, this result for \macc\, must be caused by the mass of the objects.
Indeed, we note that most of the Cha sources with higher \pab\, fluxes actually have stellar masses ($\sim$1-2~\msun) larger
than those typical of the Taurus sample (around 0.4 \msun) \citep[see e.g.][]{gatti06}).

In general, the higher \pab\, fluxes in Cha could result if the emitting surface is larger (in the optically thick LTE case), or if the temperature 
is lower (in case B scenario), or if the \mloss\, is larger (for an LTE wind) in Cha than in most Taurus CTTSs or Ophiuchus BDs.

Regarding the origin of the \hi\, emission in sources showing low $R$ values,
\citet{gatti06} tentatively interpret this result in terms of optically thick line emission originating in the accretion spots on the protostellar photosphere: 
after the mass accretion rate has 
decreased below a given threshold value, the NIR \hi\, optically thick emission component coming from the accretion spots would progressively become dominant with 
respect to the emission originating in the accretion columns. 
This interpretation appears problematic in our case, considered the emitting region sizes we derive from our observations
(the reader is warned that the three blackbody emission curves in Fig.~7 of \citet{gatti06} refer, respectively, to emitting areas ten times smaller than ours),
which would correspond to areas larger than the typical stellar disc.
In any case, in this scenario we would expect a correlation between the observed low $R$ ratios and the mass accretion rate \macc.

To investigate this aspect, we analyse the ratio Pa$\beta$/Br$\gamma$ as a function of \macc\, in Fig.~\ref{fig:ratios2}. This 
can be compared to Fig.~5 of \citet{gatti06}, which shows results for Ophiuchus BDs and for Taurus-Auriga CTTSs \citep[][also reported in our figure]{muzerolle98a}. 
In our sample, low $R$ values do not seem to be linked to a particular range of mass accretion rates, because they are observed in sources spanning a wide range 
of \macc\, (3 dex), with no evident correlation between the two quantities. 
Conversely, there is a slight indication in our sample that the highest ratios ($R>4$) are observed only 
in objects characterised by the lowest values of \macc, whereas high ratios are associated also with quite high mass accretion 
rates ($\sim$10$^{-8}$-10$^{-6}$ \msunyr) in the Taurus CTTSs of \citet{muzerolle98a}. 
To understand whether low $R$ sources possess any other common properties, we also looked for correlations between $R$ and other 
stellar properties. However, we find that $R$ does not clearly correlate with either the spectral index $\alpha_{2-24}$, \lstar, \mstar, or $T_\mathrm{eff}$.

A viable scenario to explain the observed low ratios $R$ is
the presence of an outflow, possibly associated with geometrical effects. For instance, the high-density
regions at the base of a collimated jet seen almost face-on could likely be characterised by compact
optically thick emission, where only the sections far from the star at lower temperature are observed.
Alternatively, the \hi\, emission might arise in the inner region of an optically thick gaseous
disc. However, appropriate models for such a scenario have not been developed yet, because only the
optically thin emission from the dust-free gaseous discs of relatively massive Herbig stars have
been considered so far \citep[e.g.][]{muzerolle04}.

\section{Conclusions}
\label{sec:conclusions}

As part of the POISSON project, we have presented the results of low-resolution spectroscopic optical-infrared observations (0.6-2.4 \um) 
of 47 YSOs (mostly Class II objects) in the Chamaeleon I and II star-forming regions.
The spectra, acquired with EFOSC2 and SofI at ESO/NTT, show many emission 
features commonly observed in young objects, in particular
\hi\, lines, \caii, and \oi\, emission.
Taking advantage of the wide range of wavelengths 
covered by our data, we have considered five different accretion tracers
(\brg, \pab, \ha, \caii\,$\lambda8542$, \oi\,$\lambda6300$) 
to measure the accretion luminosity \lacc, making use of existing empirical 
relationships between emission line luminosity and \lacc. 
Our analysis primarily focussed on comparing \lacc\, determinations from the five tracers, 
thereby allowing us to discuss the reliability 
and consistency of the different empirical relationships considered.
From the derived accretion luminosities we have finally computed the mass accretion rate of the objects.

The main conclusions of our work can be summarised as follows:

\begin{itemize}

\item The different tracers provide \lacc\, values characterised by different scatters when plotted as a function of \lstar.
The \brg\, relationship \citep{calvet04} appears to be the most reliable, because 
it gives the minimum dispersion of \lacc\, throughout the range of \lstar, whereas the other tracers, especially \ha, provide much more 
scattered \lacc\, results that are not expected for the homogeneous sample of targets observed. 
We therefore adopted the accretion luminosities computed from \brg.

\item The direct comparison between \lacc(\brg) and the accretion luminosity obtained from the other four tracers shows that the empirical relationships 
provide mean accretion luminosities that are systematically different in given ranges of \lstar. 
These discrepancies as well as the large scatters may probably be ascribed to different excitation mechanisms (and/or emitting regions) contributing to the line, 
which may vary between our sample and those where the relationships were calibrated, which are basically composed of Taurus objects. 
Our analysis shows that the relationships for \ha, \oi, and \caii\, are most subjected to these biases when applied to typical YSO samples. 

\item The derived accretion luminosities are in the range 0.1~\lstar-1~\lstar for all sources.
The inferred mass accretion rates span from $10^{-7}$ to $10^{-9}$ \msunyr,
in the range of values commonly observed in Class IIs. 
The \macc values computed in Cha I are roughly proportional to \mstar$^2$, in agreement with the results found
in other low-mass star-forming regions. Besides this dependence, we obtained indications of a slightly lower mean \macc\, 
(for a given \mstar) in Cha II than in Cha I. This might reflect the different age of the two associations, thus 
suggesting a decrease of the accretion rate with time.

\item We found that for \lacc(\brg) and \lacc(\pab) the observed discrepancies can be actually related to different intrinsic \pab/\brg\, ratios, 
which point to the existence of two different \hi\, emission modalities. 
\\
The first one seems to agree generally with predictions of both wind and accretion models.
Applying a simple spherical wind model, we obtained indications of gas temperatures around 4000~K and \mloss\, of $10^{-8}$-$10^{-9}$ \msunyr.
\\
The second one, associated with particularly low \pab/\brg\, values ($R<2.5$), indicates optically thick emission from regions with a size 
of the order of 10$^{22}$~cm$^{-3}$ (i.e. larger than the typical stellar surface area of an YSO) and low gas temperatures $T\lesssim4000$~K. 
Possible scenarios include emission at the base of a jet, from an inner gaseous disc, or 
from accretion-heated spots on the stellar surface (at least in the case of smallest emission regions). 
Additional investigations are definitively required to ascertain the nature of the two different emission mechanisms observed.

\end{itemize}

\section*{Acknowledgements}
\begin{small}
The authors are grateful to K.L. Luhman for providing mass estimates for the Cha I objects.
Alessio Caratti o Garatti acknowledges support from the Science Foundation of Ireland, grant
07/RFP/PHYF790, and the European Commision, grant ERG249157.
\end{small}

\bibliographystyle{aa} 
\bibliography{Simo_Ref_01} 

\section*{Appendix: Empirical line-accretion relationships}
\label{sec:appendix:lines}

We report here the empirical line luminosity-\lacc\, relationships 
for the tracers considered in our analysis, namely \oi\,$\lambda$6300, \caii\,$\lambda$8542, and the \hi\, lines \ha, \pab, and \brg,
providing a short overview of the procedures used to derive the formulae. 
More details can be retrieved in the articles where the single relationships are provided.

In general, the reliability of a given relationship will depend on how accurately the accretion luminosity 
has been determined as well as on the range of stellar masses investigated. 
UV/blue-band excess measurements are the best direct accretion diagnostics \citep[e.g.][]{hartmann98,gullbring98,gullbring00,herczeg08},
although their use is often limited owing to high extinctions typical of star-forming clouds. \lacc\, values based on veiling estimates in
other bands usually suffer from larger uncertainties \citep[e.g.][]{herczeg08}.
In this respect we note that the five relations considered are all initially based on observations of young sources of 
the Taurus-Auriga complex, for which these UV/blue-band measurements are available.
\\
\\
\noindent
\textbf{\oi}. Forbidden emission of \oi\,$\lambda$6300 is believed to arise in winds/jets powered by accreting sources 
\citep[e.g.][]{cabrit90,hamann94,hartigan95}, so that it is expected to correlate with \lacc, although in an indirect way. 
A relation between \oi\,$\lambda$6300 luminosity and accretion luminosity has been derived by \citet{herczeg08} from observations of a 
sample of young low-mass stars and brown dwarfs in Taurus, integrating their data with the results of spectroscopic measurements 
of CTTSs in the same region by \citet{hartigan03}, to cover a range of stellar masses 0.05-1 \msun.
\citet{herczeg08} derive accretion luminosities from direct measurement of the blue continuum excess, whereas \citet{hartigan03} provide
\lacc\, based on veiling in the $R$ band.

\begin{equation}
\mathrm{Log}\;L_{acc}/L_\odot = 0.96 \cdot \mathrm{Log}\;L_{\ion{[O]}{i}}/L_\odot + 2.80
\end{equation}
\textbf{\caii}. The \caii\, infrared triplet ($\lambda$8498 ,$\lambda$8542, $\lambda$8662) is believed to originate from gas in the magnetospheric 
accretion region, because the flux, in particular that of the $\lambda$8542 line, appears to correlate strongly with the accretion rate \citep{muzerolle98a}. 
However, these lines have been detected also in jets \citep{nisini05b,podio06},
which indicates that \caii\, might be (at least in part) an indirect tracer.
We adopt the \lacc-\caii $\lambda$8542 relationship computed by \citet{dahm08} on the basis of the accretion luminosities of 14 Taurus-Auriga sources,
established either through $U$-band photometry or blue excess measurement, mostly taken from \citet{gullbring98} and \citep{muzerolle98a}, 
which have masses $\lesssim$1\msun.
\citet{dahm08} then applied this relationship to his IC348 cluster sample, which displays a range of masses (0.5-2.0 \msun) similar to that of our sample:
\begin{equation}
\mathrm{Log}\;L_{acc}/L_\odot = 1.02 \cdot \mathrm{Log}\;L_{\ion{Ca}{ii}}/L_\odot + 2.50 ~. 
\end{equation}
\textbf{\ha}. \ha\, equivalent width has been traditionally used in the past to distinguish between the so-called classical and weak T Tauri stars, that is,
between actively accreting T Tauri stars and those that are relatively devoid of circumstellar matter . 
The \ha\, velocity width at 10\% of the peak has been found to be a good diagnostic of the accretion 
\citep[e.g][]{white03,natta04}
As for the flux of the line, the empirical correlation between accretion and \ha\, has been recently revised by
\citet{fang09}, who have combined the available data from \citet{gullbring98,dahm08,herczeg08}
to derive the following relationship:
\begin{equation}
\mathrm{Log}\;L_{acc}/L_\odot = 1.25 \cdot \mathrm{Log}\;L_{\mathrm{H\alpha}}/L_\odot + 2.27 ~.
\end{equation}
\textbf{\pab} and \textbf{\brg}. The correlation of the \pab\, and \brg\, lines with the accretion luminosity has been investigated in several works 
during the last years 
\citep[see e.g.][]{muzerolle98b,calvet00,calvet04,natta04,nisini05a}.
These correlations are particularly important because they provide a straightforward method to compute \lacc\, for young
embedded sources that are observable only at infrared wavelengths, and for which it is consequently impossible to measure accretion directly through the detection of the UV-optical veiling.
\hi\, infrared emission has therefore been extensively used as tracer to derive the accretion rate in young objects 
\citep[see e.g.][]{natta06,garcia_lopez06,antoniucci08} 
For our sample we adopt the empirical laws obtained by \citet{calvet00,calvet04}, based on accretion luminosities derived from the UV excess 
observed in a sample of Taurus and Orion objects. This extends the relations found by \citet{muzerolle98a}
for low-mass CTTSs to more massive pre-main sequence objects (up to 4 \msun).
\begin{equation}
\mathrm{Log}\;L_{acc}/L_\odot = 1.03 \cdot \mathrm{Log}\;L_{\mathrm{Pa\beta}}/L_\odot + 2.8 
\end{equation}
\begin{equation}
\mathrm{Log}\;L_{acc}/L_\odot = 0.9 \cdot \mathrm{Log}\;L_{\mathrm{Br\gamma}}/L_\odot + 2.9 
\end{equation}
%


%
\end{document}